\documentclass[pra,nofootinbib,reprint]{revtex4-2}

\usepackage{amsmath}
\usepackage{amsfonts}
\usepackage{color}
\usepackage{comment}
\usepackage{gensymb} 
\usepackage{graphicx}
\usepackage{wasysym} 
\usepackage{xspace}

\newcommand{\ie}{{i.e.,}\xspace}
\newcommand{\eg}{{\em eg}\xspace}

\newcommand{\dd}{\ensuremath{{d}}\xspace}
\newcommand{\dtc}{\ensuremath{_{\text{\tiny d}}}}

\newcommand{\oned}{\ensuremath{\text{1D}}\xspace}

\newcommand{\thetak}{\ensuremath{\theta_{\mbox{\tiny k}}}\xspace}

\newcommand{\xini}{\ensuremath{\langle x\rangle_0}\xspace}

\newcommand{\acronym}[1]{{\text{\uppercase #1}}\xspace}
\renewcommand{\acronym}[1]{{\text{\MakeUppercase{#1}}}\xspace}
\newcommand{\rabbit}{\acronym{rabbit}}
\newcommand{\RABBIT}{\rabbit}
\newcommand{\tdse}{\acronym{tdse}}

\newcommand{\tise}{\acronym{tise}}

\newcommand{\xuv}{\acronym{xuv}}
\newcommand{\XUV}{\xuv}
\newcommand{\ir}{\acronym{ir}}

\newcommand{\scwf}{\ensuremath{\text{\acronym{scwf}}}\xspace}

\newcommand{\bo}{\acronym{bo}}
\newcommand{\BO}{\bo}
\newcommand{\fc}{\acronym{fc}}
\newcommand{\FC}{\fc}
\newcommand{\fwhm}{\acronym{fwhm}}

\newcommand{\rdm}{\acronym{rdm}}
\newcommand{\RDM}{\rdm}

\newcommand{\ratio}{\ensuremath{\mathcal{R}}\xspace}
\newcommand{\ratiorab}{\ensuremath{\mathcal{R}_{\text{\tiny rab}}}\xspace}
\newcommand{\tratiorab}{\ensuremath{\tilde{\mathcal{R}}_{\text{\tiny rab}}}\xspace}

\newcommand{\wX}{\ensuremath{\omega_{\text{\small xuv}}}\xspace}
\newcommand{\tauxuvir}{\ensuremath{\tau_{\text{\small xuv-ir}}}\xspace}
\newcommand{\wIR}{\ensuremath{\omega_0}\xspace}
\newcommand{\lIR}{\ensuremath{\lambda_0}\xspace}
\newcommand{\TIR}{\ensuremath{T_0}\xspace}

\newcommand{\HA}[1]{\ensuremath{\mbox{HA}_{#1}}\xspace}
\newcommand{\SB}[1]{\ensuremath{\mbox{SB}_{#1}}\xspace}

\newcommand{\Ip}{\ensuremath{E_\acronym{i}}\xspace}

\newcommand{\Vne}{\ensuremath{V_{\text{\tiny N-e}}}\xspace}
\newcommand{\Vnn}{\ensuremath{V_{\text{\tiny N-N}}}\xspace}
\newcommand{\Hel}{\ensuremath{H_{\text{\tiny e}}}\xspace}
\newcommand{\Hnu}{\ensuremath{H_{\text{\tiny N}}}\xspace}
\newcommand{\Req}{\ensuremath{R_{\text{\tiny eq}}}\xspace}
\newcommand{\Ropt}{\ensuremath{R_{\text{\tiny opt}}}\xspace}

\newcommand{\Rvrt}{\ensuremath{R_{\text{\tiny vrt}}}\xspace}
\newcommand{\modmol}[1]{\ensuremath{\mathcal{#1}}\xspace}
\newcommand{\molA}{\modmol{A}}
\newcommand{\molB}{\modmol{C}}
\newcommand{\molC}{\modmol{B}}
\newcommand{\rdt}{\ensuremath{r\dtc}\xspace}
\newcommand{\tof}{\acronym{tof}}
\newcommand{\tdv}{\ensuremath{\tau}\xspace}
\newcommand{\Yld}{\ensuremath{\mathcal{Y}}\xspace}

\newcommand{\wigdel}{\ensuremath{\tau_{\mbox{\tiny \sc w}}}\xspace}
\newcommand{\rabdel}{\ensuremath{\tau_{\mbox{\tiny mol}}}\xspace}

\newcommand{\ion}{\ensuremath{_{\text{\tiny ion}}}}
\newcommand{\E}{\ensuremath{\mathcal{E}}\xspace}

\newcommand{\tda}{\ensuremath{a}\xspace}
\newcommand{\trace}[1]{\ensuremath{\mbox{tr}(#1)}\xspace}

\definecolor{jcol}{rgb}{1,0,0}
\definecolor{mcol}{rgb}{0.75,0.0,0.75}

\newcommand{\jref}[1]{(\ref{#1})}

\definecolor{done}{rgb}{0,0.8,0}
\definecolor{tobedone}{rgb}{0.8,0,0}

\newcommand{\figOK}{\jccomment{Figure OK}}
\renewcommand{\figOK}{}

\newlength{\figwidth}

\begin{document}
\setlength{\figwidth}{0.99\linewidth}

\title{Anisotropic molecular photoemission dynamics: Interpreting and accounting for the nuclear motion}
\author{Antoine Desrier}
\author{Morgan Berkane}
\author{Camille L\'ev\^eque}
\author{Richard Ta\"\i eb}
\author{J\'er\'emie Caillat}
\email{jeremie.caillat@sorbonne-universite.fr}
\affiliation{Sorbonne Universit\'e, CNRS, Laboratoire de Chimie Physique-Mati\`ere et Rayonnement, LCPMR, F-75005 Paris, France}
\date{\today}

\begin{abstract}
We investigate how vibration affects molecular photoemission dynamics, through simulations on two-dimension asymmetric model molecules including the electronic and  nuclear motions in a fully correlated way. We show that a slight anisotropy in the electron-ion momentum sharing is sufficient to prevent one from unambigously characterizing the vibrationnaly averaged  photoemission dynamics in terms of stereo Wigner delays. We further show that vibrational resolution can be retrieved in fixed-nuclei simulations, using effective molecular conformations that are specific to each vibrational channel. The optimal internuclear distances found empirically in 1-photon processes can be identified {\em a priori} using simple physical arguments. They also turn out to be efficient to simulate vibrationnally-resolved \rabbit measurements and to account for interchannel coherences in broadband 1-photon ionization.
\end{abstract}
\maketitle
\section{Introduction}
From its early emergence as an applied branch of attosecond sciences to its most recent and ongoing developments, attochemistry offers and envisions unprecedented ways of exploring fundamental dynamics occurring in molecules on their natural time scale~\cite{baker2006a,salieres2012a,lepine2014a,bag2017a,nisoli2017a,merritt2021a,calegari2023a}. 
In the broader perspective of attosecond time-resolved spectroscopies, photoemission endorses a singular role, either as a characterization tool~\cite{paul2001a,hentschel2001a} or as the probed process itself~\cite{cavalieri2007a,schultze2010a,klunder2011a}.  

Revisiting a process as essential as molecular photoemission in the time domain is made possible only through intertwined, long term interactions between theory and experiments, that tackle and exploit the complexities and richness specific to molecular systems both in the conception and in the interpretation of simulations or measurements. It was pioneered by investigating experimentally the intricate dynamics of resonant ionization in N$_2$ with vibrational resolution~\cite{haessler2009a}, using the `reconstruction of attosecond beatings by interferences of two-photon transitions' (\rabbit) scheme~\cite{paul2001a,muller2002a}. The interpretation of the measured channel-selective \rabbit phases in terms of transition delays was subsequently clarified by means of simulations on low-dimensional model molecules~\cite{caillat2011a,vacher2017a}. The experimental results were furthermore reproduced and expanded recently in elaborate simulations aiming at a quantitative agreement~\cite{borras2023a}. The anisotropy of ionization dynamics, which is another essential feature of molecular photoemission, was considered first in simulations of attosecond `streaking' measurements~\cite{hentschel2001a,yakovlev2010a}  on a model CO molecule~\cite{chacon2014a} and then explored experimentally, also on CO, using \rabbit~\cite{vos2018a}. Further experimental and theoretical investigations addressing anisotropy have since then been reported on systems of increasing complexities~\cite{barillot2015a,hockett2016a,baykusheva2017a,beaulieu2017a,ahmadi2022a,boyer2023a}, see also~\cite{berkane2024a} and references therein. 

The imprints of nuclear motion on photemission dynamics in polyatomic molecules is the subject of an increasing number of  theoretical and experimental investigations. Among the most recent studies, the authors of Ref.~\cite{patchkovskii2023a} investigate theoretically how nuclear motion affects photoelectron spectra in the context of \rabbit measurements. In this context, they establish an expression of molecular \rabbit spectra in terms of a convolution between a nuclear autocorrelation function and two-photon electronic transition matrix elements, following an approach initially derived to simulate molecular streaking~\cite{kowalewski2016a}. In \cite{gong2023a}, the imprints of nuclear motion on attosecond photoemission delays is investigated experimentally on the methane molecule and its deutered counterpart, using \rabbit and an advanced theoretical support. The authors find no significant isotopic effect, but a $\sim 20$ as difference between dissociative and non-dissociative channels. A closely related study on the same polyatomic molecules further  highlighted how nuclear motion impacts the coherence of the photoemission process~\cite{ertel2023a}, which is an essential issue for most of the interferometric attosecond resolved pump-probe schemes.

 Nonetheless, investigations on smaller -- diatomic --  benchmark molecules~\cite{cattaneo2018a,bello2018a,nandi2020a,vrakking2021a,wang2021a,liao2021a,vrakking2022a,gong2022a,chen2023a,serov2023a,li2023a,ke2023a,nabekawa2023a} remain of great importance to gradually and comprehensively explore the interplay of the molecular degrees of freedom from a time-dependent perspective.  On the theory side, simulations on simple models with limited degrees of freedom~\cite{caillat2018a} are essential to guide and interpret experiments and elaborate simulations in terms of intuitive pictures and practical notions.

\begin{figure*}
\includegraphics[width=2\figwidth]{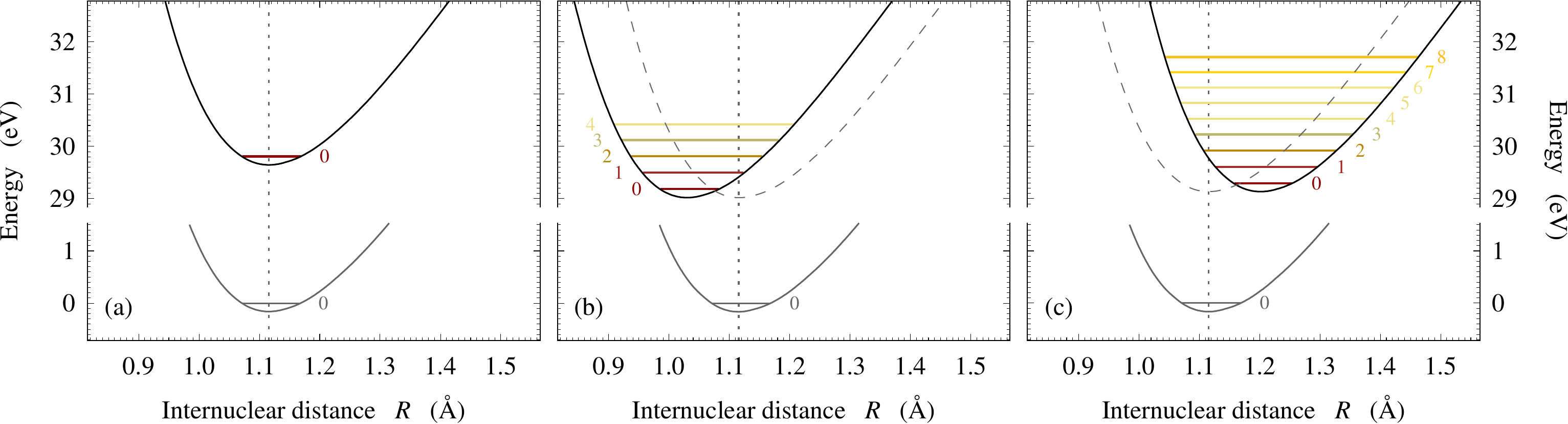}
\caption{\label{fig:BOcurves} (Color online) Born-Oppenheimer energies of the model molecules \molA [frame (a)], \molC [frame (b)] and \molB [frame (c)] designed for the simulations, as a function of the internuclear distance $R$. The energy scale refers here to the overall ground states of the neutral molecules. For each molecule, the full  grey and black curves correspond to the neutral and ionic electronic ground states, respectively. In frames (b) and (c), the dashed grey curve is an image of the neutral ground curve vertically shifted towards the ionization threshold, displayed for comparison purpose. The vertical dotted lines indicate the equilibrium distance of the neutral \Req. The vibrational levels $v$ presenting a significant overlap with the ground state [$\vert F(v)\vert^2>10^{-2}$, see Eq.~\jref{eqn:FC} and Fig.~\ref{fig:FC}] are indicated in the ionic curve of each molecule \figOK. 
}
\end{figure*}

In the present article, we investigate numerically how vibration imprints the attosecond dynamics of anisotropic molecular photoemission even when the vibronic couplings are minimal -- notably far from any vibronic resonance\footnote{It comes as a companion paper to Ref.~\cite{berkane2024a}, which is focused on the definition and \RABBIT measurement of anisotropic molecular photoemission delays, regardless of the nuclear motion.}. Our study is based on  low dimensional model molecules allowing for extensive numerically exact simulations. The objectives are two-fold: i) to study the interplay of molecular asymmetry and nuclear motion on attosecond time-resolved photoemission, and ii) to identify a way to account for the vibrational resolution in standard theoretical  approaches with fixed nuclei -- keeping in mind that only the latter can be routinely applied to the simulations of more realistic models.

The paper is organized as follows. Section~\ref{sec:toolbox} presents the model molecules and the overall methodology followed in our study. In section~\ref{sec:A1ph}, we benchmark the signatures of molecular asymmetry on 1-photon ionization dynamics with a model  displaying no effective nuclear motion during and upon photoemission. It partly reproduces and expands the results presented and commented in~\cite{berkane2024a}. In section~\ref{sec:BC1ph}, we consider photoemission occurring with sensible nuclear motion. We question the relevance of defining an orientation-resolved photoemission delay for the photoelectron wave packet averaged over the open vibrationnal channels. Then we consider vibrationnaly-resolved photoemission dynamics, and investigate the possibility to reproduce it in a fixed-nuclei approach. In Section~\ref{sec:applications} we address the capacities of the fixed nuclei approach to reproduce coherent, vibrationnally-resolved, photoemission beyond the context in which it was designed. The conclusions are presented in section~\ref{sec:conclusion}.

Unless stated otherwise, the equations are displayed in atomic units (a.u.) all through the paper.

\section{Theoretical toolbox}\label{sec:toolbox}

	\subsection{Model molecules}\label{sec:modmol}

Our simulations were performed on single-active electron diatomic model molecules including correlated electronic and internuclear motions each in \oned. They are similar to the ones used \eg in \cite{lein2005a,houfek2006a,caillat2011a}. 
\subsubsection{Generic hamiltonian}
The generic field-free hamiltonian of the model molecules reads
\begin{eqnarray}\label{eqn:H0}
H_0&=&\underbrace{-\frac{1}{2\mu}\frac{\partial^2}{\partial R^2}+\Vnn(R)}_{\Hnu}\underbrace{-\frac{1}{2}\frac{\partial^2}{\partial x^2}+\Vne(x,R)}_{\Hel}
\end{eqnarray}
where the two coordinates $x$ and $R$ are the electron position and the internuclear distance, respectively, $\mu$ is the nuclei reduced mass, $\Vnn(R)$ is the interatomic potential in absence of the active electron, and $\Vne(x,R)$ is the interaction potential between the active electron and the molecular ion. 

The interatomic potential \Vnn for each molecule is defined numerically on a discretized $R$-grid, while the electron-nuclei potential \Vne is defined as an asymmetric two-center soft-Coulomb potential,
\begin{eqnarray}\label{eqn:Vne}
\Vne(x,R)&=&\sum\limits_{j=1,2}-\frac{q_j}{\sqrt{(x-X_j)^2+a(R)}}
\end{eqnarray} 
where
\begin{eqnarray}\label{eqn:Xj}
X_j&=&(-1)^j\frac{\mu}{M_j}R
\end{eqnarray}
 are the positions of each nucleus ($j=1,2$), of mass $M_j$, with respect to their center of mass. The \Vne potential is adjustable through the effective charges $q_j>0$ (with the constraint $q_1+q_2=1$) and the screening parameter $a(R)>0$ defined numerically over the same $R$-grid as \Vnn.  One should note that \Vne accounts not only for the asymmetry related to the electronegativity difference (through $q_j$) but also for the mass asymmetry (through $X_j$). 
 
 In practice, we adjusted \Vnn and \Vne  empirically~\cite{caillat2018a} for the energies of the model molecule to match some desired energy curves within the Born-Oppenheimer (\BO) approximation: \Vnn is directly the molecular ion energy curve, while the negative eigen-energies of $\Hel+\Vnn$ computed at each $R$ [see partition of $H_0$ in Eq.~\jref{eqn:H0}] provide the electronic ground and excited curves of the neutral. Within the \bo framework, the ground state energy $\varepsilon_0(R)$ of the electronic hamiltonian \Hel  is the opposite of the vertical ionization potential for each $R$,
\begin{eqnarray}\label{eqn:IpR}
\tilde\Ip(R)&=&-\varepsilon_0(R),
\end{eqnarray}
\ie the energy gap between the ionic and the neutral ground state curves.
 
 \subsubsection{Model parameters}
We considered three model molecules, hereafter referred to as \molA, \molC and \molB for simplicity. For each of them, the effective charges were set to $q_1=0.33$ a.u. and $q_2=0.67$ a.u, and the nuclei masses to $M_1=20.53$~u and $M_2=7.47$~u. The chosen masses are such that $M_1+M_2=28$~u, \ie the mass of CO -- a prototypical hereronuclear diatomics on which orientation- and time-resolved photoemission simulations~\cite{chacon2014a} and experiments~\cite{vos2018a} have already been performed. However, our models are only loosely based on CO, as we, among other liberties, increased the mass ratio $M_2/M_1$ in order to emphasize the consequences of mass asymmetry on photoemission dynamics.  The potentials were adjusted to obtain the \BO curves displayed in Fig.~\ref{fig:BOcurves}. 
The three molecular systems share the same ground state energy curve, which mimics the one of CO~\cite{krupenie1965a}, with an equilibrium distance $\Req=1.115$~\AA. 

In molecule \molA, the ionic curve is a vertically shifted image of the ground state. In molecule \molC (resp. \molB), it is also a replica of the ground state curve, translated towards lower (resp. larger) values of $R$ with its minimum at $1.029$~\AA\ (resp. $1.201$~\AA). 
 The energy shifts of the ionic curves were adjusted to achieve $\tilde\Ip(\Req)\simeq 30$ eV for all molecules ($29.81$ eV,  $29.18$ eV and $29.30$ eV for molecules \molA, \molC and \molB, respectively). As an illustration, the $x$-dependency of the electron-core potential for molecule \molA at the equilibrium distance $R=\Req$ can be found in Fig.~\ref{fig:wfnA}(a).  

Due to the relative positions of these energy curves, photoionization takes place with no effective nuclear motion in molecule~\molA, while it initiates a bond contraction in molecule \molC and an elongation in molecule~\molB.

	\subsection{Methodology overview}
We considered two different approaches for the simulations. The first one consists in simulating the `complete' vibronic dynamics of the model molecules. To this end, we performed numerically converged, vibronically correlated, simulations based on the resolution of the time-dependent Schr\"odinger equation (\tdse). 
The second approach consists in adopting a standard simplified treatment with fixed nuclei, based on the Born-Oppenheimer (\bo) approximation, where the photoemission dynamics are encoded in the continuum solutions of the electronic time-independent Schr\"odinger equation (\tise). 

Our motivation is two-fold. On the one hand, the `complete' vibronic simulations highlight the impact of vibronic couplings on the dynamics of molecular photoemission investigated with attosecond resolution. On the other hand, by confronting these `exact' dynamics to the ones obtained in the \bo framework, we assess the relevance of fixed-nuclei approaches to investigate molecular photoemission process (see \eg~\cite{gianturco1994a,natalense1999a,chacon2014a,huppert2016a,baykusheva2017a,chacon2018a}), when the latter are performed with well-chosen molecular conformations. In all the simulations, the $x$ origin was set \textit{a priori} to coincide with the center of mass of the molecule, consistently with the definition of the \Vne potential [see Eqs.~\jref{eqn:Vne} and~\jref{eqn:Xj}]. 

The central quantities used in this work to characterize anisotropic photoionization dynamics are orientation-resolved yields and  stereo ionization delays, which can be computed indifferently in both approaches.

\subsubsection{`Complete' simulations}	
The complete dynamics of the molecules were simulated by solving numerically the vibronic \tdse
\begin{eqnarray}\label{eqn:tdse}
i\frac{\partial\Phi(x,R,t)}{\partial t}&=&[H_0+W(t)]\Phi(x,R,t).
\end{eqnarray}
Here $\Phi(x,R,t)$ is the propagated vibronic wave-function and $W(t)$ represents the dipole interaction of the molecule with the ionizing pulse, implemented in the velocity gauge. The initial state is the ground vibronic state $\Phi_0(x,R)$, obtained by imaginary time propagation, in all the simulations. Numerically exact resolutions of the \tdse [Eq.~\jref{eqn:tdse}] were performed using a split operator propagation based on a grid representation of the $x$ coordinate, combined with an expansion over a set of eigenvectors of \Hnu for the $R$ coordinate, as detailed in~\cite{caillat2011a,caillat2018a}. 
The vector potentials of the ionizing pulses were assigned $\sin^2$ temporal envelops with central photon energies \wX corresponding to harmonics of a Ti:Sapphire laser ($\lIR=800$ nm, $\wIR=1.55$ eV) in the extreme ultraviolet (\xuv).  Case specific pulse parameters are indicated further in the text. Simulations on molecule \molA were performed by expanding the wave-function on a single vibrationnal state ($v=0$ is the only open channel). For molecules \molC and \molB, the results shown in this paper were obtained by including all vibrational channels up to $v=8$ and $17$, respectively, which safely ensures convergence.

The analysis of these complete time-dependent simulations are mostly based on the outgoing electron flux computed in the asymptotic $x$ region, where the short-range components of \Vne (including the vibronic couplings) vanish. 

\subsubsection{Simulations at fixed internuclear distances}
In the \bo framework, we treated 1-photon ionization by analyzing electronic continuum wave-functions computed at fixed internuclear distances. Among the solutions of the \tise
\begin{eqnarray}\label{eqn:tise}
\Hel\psi_\varepsilon(x)&=&\varepsilon\psi_\varepsilon(x)
\end{eqnarray}
in the degenerate continuum (at energies $\varepsilon>0$), we worked with the  wave-functions specifically selected by 1-photon transitions starting from the ground state, hereafter referred to as  \scwf for `selected continuum wave-functions'. These real-valued continuum wave-functions are defined unambiguously, and their analysis and interpretation 
are 
independant from their definition and computation. The \scwf formalism was introduced in~\cite{gaillac2016a} and later used in~\cite{liao2021a,berkane2024a}.  Note that the electronic hamiltonian \Hel [as defined in Eq.~\jref{eqn:H0}], its eigenfunctions $\psi_\varepsilon(x)$ and eigenvalues $\varepsilon$ depend parametrically on the internuclear distance $R$ -- although it does not appear explicitely in Eq.~\jref{eqn:tise} for the sake of readability.

Within this  framework, the dynamics of photoemission are encoded in the asymptotic features of the \scwf. 
\\ \\
 More computational details are provided along with the presentation of the results. 

	\section{Photoemission from molecule \molA: signatures of the potential asymmetry}\label{sec:A1ph}
	
We present here the results obtained for molecule~\molA, which was designed to benchmark our methodological approach. We first detail the analysis of the time-dependent vibronic simulations, and then verify that the latter are consistent with time-independent \bo simulations at equilibrium distance.

		\subsection{`Complete' simulations}\label{sec:A1phcomp}
		
In these simulations, the XUV pulse intensities were set to $10^{12}$ W/cm\textsuperscript{2}, \ie low enough to avoid any significant multi-photon process in 1-color photoionization. The total pulse durations were set to $15.87$ fs, which corresponds to 6 fundamental periods ($\TIR=2\pi/\wIR=2.67$ fs). 

			\subsubsection{Photoelectron flux}
			
\begin{figure}[t]
\includegraphics[width=\figwidth]{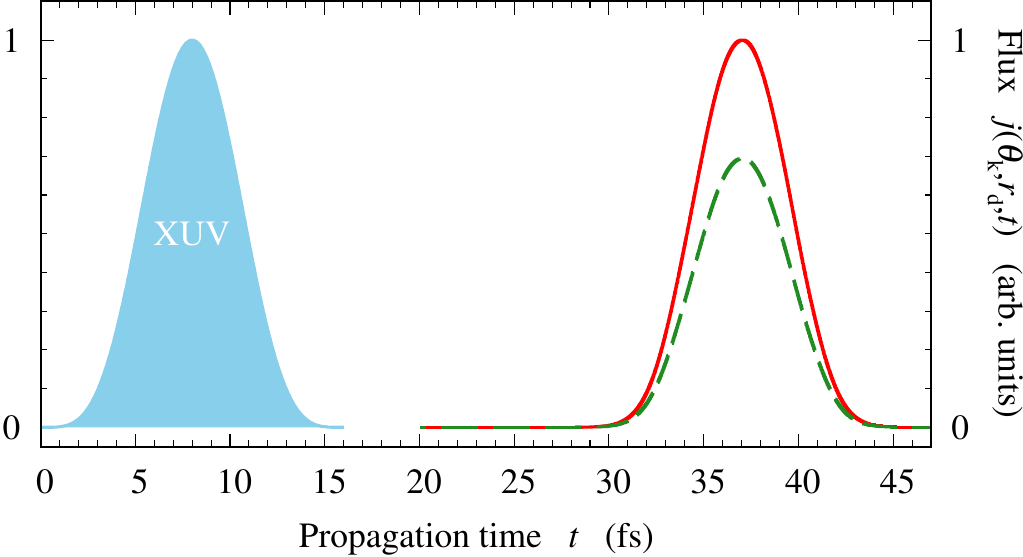}
\caption{\label{fig:flxA} Outgoing electron flux $j(\thetak,\rdt,t)$ computed at $\rdt=800$ a.u. on the left ($\thetak=180\degree$, red full curve) and right ($\thetak=0\degree$, green dashed curve) sides of molecule \molA ionized by a $\wX=35.67$ eV pulse, as a function of propagation time $t$. The flux are normalized to 1 at the overall maximum  reached on the left side. The temporal envelop of the ionizing pulse is represented by the blue filled curve.  
}
\end{figure}
In the complete time-dependent simulations, we characterized the ionization dynamics through the outgoing photoelectron flux computed at a given detection distance \rdt from the $x$-origin, on either side of the molecule and averaged over $R$:
\begin{eqnarray} \label{eqn:intj}
j(\thetak,\rdt,t)&=&\cos\thetak\times\\
 \nonumber
&&\mathfrak{Im}\int\limits_0^\infty\, 
\Phi^\star(\rdt\cos\thetak,R,t)
\Phi'(\rdt\cos\thetak,R,t)
\,\dd R 
\end{eqnarray}
where $\Phi^\star(x,R,t)$ is the complex conjugate of the  vibronic wave-function  propagated according to Eq.~\jref{eqn:tdse}, $\Phi'(x,R,t)$ its derivative with respect to $x$. Here and all through the paper, \thetak represents the direction of photoemission, restricted to two discrete values: $\thetak=0\degree$ (emission towards the right, $x>0$) and $180\degree$ (towards the left, $x<0$).

As an illustration, Figure~\ref{fig:flxA} shows the flux computed at $\rdt=800$ a.u. (423 \AA) on the left and right sides of  molecule \molA ionized by a pulse of central frequency $\wX=23\times\wIR=35.67$ eV. The flux profiles follow the ionizing pulse envelop (also shown), shifted by $\approx  30$ fs, which is consistent with the time needed for a (nearly) free electron with the energy $\wX-\tilde\Ip(\Req)=5.86$ eV to cover the distance \rdt.  The flux maximum  $\sim1.5$ times larger on the left side than on the right side is a clear signature of the photoemission anisotropy. 

		\subsubsection{Orientation-resolved yields and delays}\label{sec:molA-molbit}
\begin{figure}[t]
\includegraphics[height=1.1\figwidth]{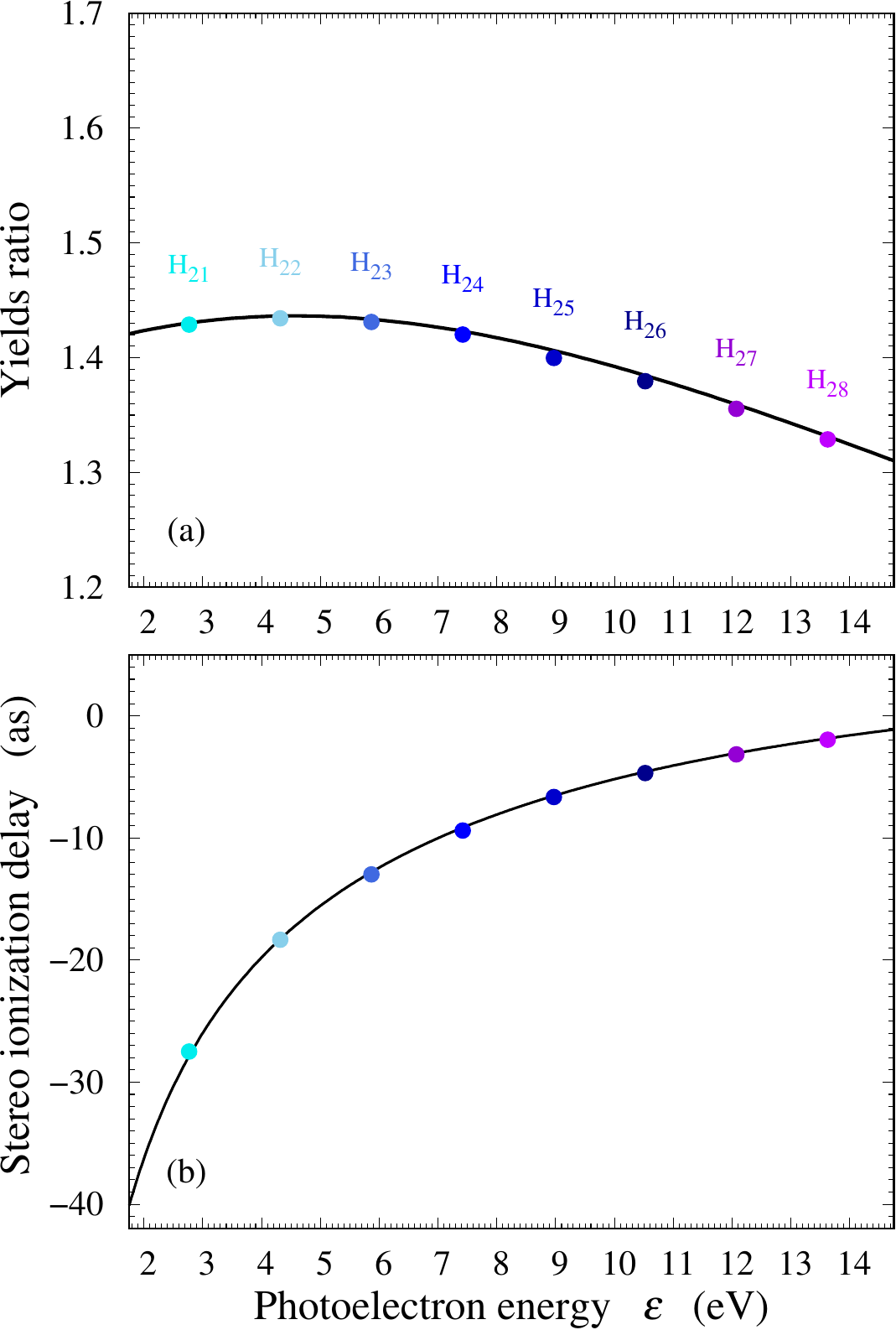}
\caption{\label{fig:ionA} Anisotropic photoemission from molecule \molA. (a) left/right ionization yields ratio computed in the fully correlated simulations [Eq~\jref{eqn:ratio}] and in the \bo framework [Eq~\jref{eqn:tratio}], resp. $\ratio$  (circles) and $\tilde\ratio$  (line); (b) stereo ionization delays computed in the fully correlated simulations [Eq~\jref{eqn:tau}] and in the \bo framework [Eq~\jref{eqn:ttau}], resp.  $\Delta\tau$ (circles) and $\Delta\wigdel$ (line). All data plotted as functions of the photoelectron energy $\varepsilon$. \figOK 
}
\end{figure}
We thus computed the orientation-resolved yields,
\begin{eqnarray}\label{eqn:tdYRL}
\Yld(\thetak)&=&\int\,j(\thetak,\rdt,t)\,\dd t,
\end{eqnarray}
in a series of simulations with field frequencies corresponding to harmonics 21 to 24, \ie \wX ranging from $32.57$ eV to $37.23$ eV by $1.55$ eV steps.
 The corresponding yields ratios
 \begin{eqnarray}\label{eqn:ratio}
 \ratio&=&\frac{\Yld(180\degree)}{\Yld(0\degree)}
 \end{eqnarray}
  are displayed in Fig.~\ref{fig:ionA}(a) (coloured circles) against the photoelectron energy $\varepsilon$. 
It evolves smoothly slightly above 1.4, indicating that photoemission is  sensitive to the asymmetry of the ionic potential, with little variations over the energy range covered in the simulations.

To characterize the asymmetry of photoemission in the time domain, we used the average ``time of flight'' (\tof)~\cite{caillat2011a,gaillac2016a,vacher2017a} towards the detector in each direction  as
\begin{eqnarray}\label{eqn:tdv}
\tdv(\thetak)&=&\frac{\int\limits\, t\times j(\thetak,\rdt,t)  \,\dd t}{\int\limits\, j(\thetak,\rdt,t)  \,\dd t}.
\end{eqnarray}
To further investigate the angular variations of the \tof computed as such, one must take into consideration the arbitrary origin $x=0$ on either sides of which the virtual detectors are located, and which does not coincide with the average initial position of the electron. In the following, we will therefore characterize these angular variations with the origin-corrected  \tof difference,
\begin{eqnarray}\label{eqn:tau}
\Delta\tau&=&\tdv(180\degree)-\tdv(0\degree)-2\frac{\xini}{\sqrt{2\varepsilon}},
\end{eqnarray}
where \xini is the average initial electron position in the ground state of the molecule $\Phi_0(x,R)$,
\begin{eqnarray}\label{eqn:x0}
\xini=\langle \Phi_0 \vert x \vert \Phi_0 \rangle.
\end{eqnarray}
The role of the last term  on the r.h.s of Eq.~\jref{eqn:tau} is discussed in details in the companion paper~\cite{berkane2024a} in the context of time-independent approaches. Its justifications identically holds for time-dependent simulations. It compensates a spurious shift appearing when computing ionization delays with an arbitrary origin -- while the photoelectron {\em in average} originates from \xini. We verified numerically that including this term (after computing the {\tof}s) is equivalent to shifting the potential such that $\xini=0$ (prior to solving the \tdse). 
In the present simulations, as well as in molecules \molC and \molB, $\xini=-0.160$ \AA. 

The stereo ionization delays $\Delta\tau$ obtained for the considered set of \xuv frequencies are plotted in Fig.~\ref{fig:ionA}(b). Starting from $\sim -40$ as just below $\varepsilon=2$ eV, its magnitude decays smoothly while the photon energy increases, as could be expected in absence of significantly structured continuum. Note that such attosecond delays cannot be resolved visually in Fig.~\ref{fig:flxA}, since they are extremely small compared to the temporal spread of the photoelectron wave-packet  at detection ($\sim 15$ fs, approximately the \xuv pulse duration). They are nonetheless significant and we have ensured their numerical convergence.

Among a series of standard numerical checks, we verified that the measured stereo delays $\Delta\tau$ do not depend on the detection distance \rdt (as long as it lies far enough from the ionic core), which is a fundamental property of the short-range scattering delays.

	\subsection{Simulations at fixed internuclear distances}

Here, we consider the molecule \molA at its equilibrium internuclear distance, see Fig.~\ref{fig:wfnA}(a). It corresponds to the \oned model molecule used for the numerical experiments presented and analysed in~\cite{berkane2024a}.

		\subsubsection{Selected continuum wave-function}

\begin{figure}[t]
\includegraphics[width=0.9\figwidth]{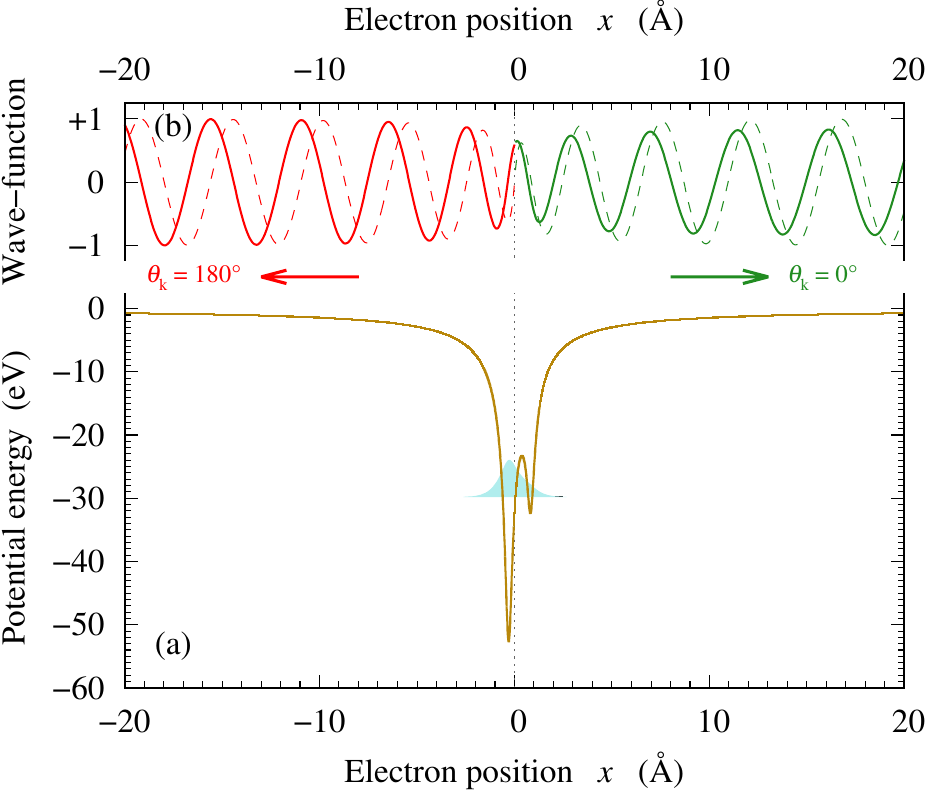}
\caption{\label{fig:wfnA} Molecule \molA in the \bo framework (equilibrium internuclear distance). (a) Electron-nuclei potential $\Vne(x,\Req)$ as a function of the electron position $x$ (dark yellow full curve). The ground state electronic wave-function is also shown (light blue filled curve). (b) Electronic continuum wave-function (full curve) selected at the energy $\varepsilon=5.86$~eV by a 1-photon transition from the electronic  ground state. Odd-parity reference wave-function (dashed curve) used to define and compute the orientation-dependent phase-shifts. The displayed continuum wave-functions are normalized such that their amplitudes asymptotically converge to 1 on the left-hand side of the molecule.
In this Figure, the left/right discrimination and the parity refer to the arbitrary $x=0$ origin (indicated by a vertical dotted line). 
}
\end{figure}

The \bo approach relies on the analysis of the electronic continuum wave-functions $\psi_\varepsilon(x)$ associated with the photoemission processes. We thus computed the \scwf of molecule \molA by solving Eq.~\jref{eqn:tise} over the same energy range as in the full time-dependent simulations, with a fixed internuclear distance set to \Req. 

The \scwf computed at the illustrative energy $\varepsilon=5.86$~eV (corresponding to $\wX=23\times\wIR=35.67$ eV, as in Fig.~\ref{fig:flxA}) is shown in Fig.~\ref{fig:wfnA}(b) (full curve). The pseudo-period of the oscillations on both sides is consistent with a kinetic energy asymptotically converging to $5.86$ eV.  Here, the anisotropy of photoemission is clearly visible in the asymmetry of the {\em amplitudes}  of the \scwf on either sides of the molecule. 

		\subsubsection{Orientation-resolved yields and delays}\label{sec:molA-scwf}

For each pulse frequency considered in the fully-fledge approach, an alternative evaluation of the orientation-dependent ionization yields is provided, up to a global factor,  as
\begin{eqnarray}\label{eqn:tiYRL}
\tilde \Yld(\thetak)&=&\vert A(\Req;\thetak)\vert^2,
\end{eqnarray}
where  $A(\Req;,\thetak)$ is the asymptotic amplitude of the \scwf computed with $R=\Req$, at the average photoelectron energy, on each side of the molecule. We evaluated these amplitudes using Str\"omgren's normalization procedure (see~\cite{seaton1965a} and references therein). 

The corresponding yields ratio
 \begin{eqnarray}\label{eqn:tratio}
 \tilde\ratio&=&\frac{\tilde \Yld(180\degree)}{\tilde \Yld(0\degree)}
 \end{eqnarray}
computed over the considered energy range is displayed as a full line in Fig.~\ref{fig:ionA}(a). The perfect agreement with the ratio $\ratio$ computed in the vibronic time-dependent simulations  [Eq.~\jref{eqn:ratio}] constitutes a first numerical validation of our comparative approach. 

From the time domain perspective, we analyzed the \scwf in terms of orientation resolved Wigner delays~\cite{wigner1955a}, as detailed in~\cite{berkane2024a}. In few words, the Wigner delay is defined as the spectral derivative
\begin{eqnarray}
\wigdel(\thetak)&=&\frac{\partial \eta(\thetak)}{\partial \varepsilon}
\end{eqnarray}
of the asymptotic phase shifts of the \scwf computed on each side of the model molecule, with respect to an arbitrary intermediate reference (here, radial Coulomb $s$-waves centered at $x=0$). As explained in~\cite{berkane2024a}, and consistently with the `complete' vibronic simulations, we characterized the angular variations of the photoemission dynamics in the \bo framework through the origin-corrected stereo Wigner delay
\begin{eqnarray}\label{eqn:ttau}
\Delta\wigdel&=&\wigdel(180\degree)-\wigdel(0\degree)-2\frac{\xini}{\sqrt{2\varepsilon}}.
\end{eqnarray}
Since the exact ground state of the molecule is accurately modelled by its \bo counterpart,  the value of the initial average position \xini is the same here as in the `complete' simulations. The stereo Wigner delays computed in the considered energy range, already shown in Fig.~3 of \cite{berkane2024a}, are reported in Fig.~\ref{fig:ionA}(b) (full line). They perfectly agree with the stereo delays computed in the `complete' time-dependent simulations. 

Apart from benchmarking our comparative approach, the results obtained with molecule \molA put forwards the signatures of the left-right electronegativity asymmetry of the molecule on the photoelectron dynamics. 
The perfect agreement observed in Fig.~\ref{fig:ionA} between the `complete' and the fixed nuclei approach comes as no surprise since  (i) the equivalence between the time-dependent and time-independant approaches has already been put forward  by Wigner when interpreting in the time-domain the group delay associated with a scattering phase shift~\cite{wigner1955a}, and (ii) photoemission of molecule \molA occurs with no effective nuclear motion, \ie the electron and nuclear degrees of freedom, $x$ and $R$, are \textit{de facto} uncoupled  due to the relative positions of the ground state and ionic curves [see Fig.~\ref{fig:BOcurves}(a)]. 

We now move on to the results obtained with the \molC and \molB  molecules, for which photoemission occurs along with sensible nuclear dynamics.

	\section{Photoemission from molecules \molC and \molB: signatures of the vibronic couplings}\label{sec:BC1ph}
	

The necessity to take the nuclear motion into consideration in near-threshold photoemission from molecules \molC and \molB is a consequence of the relative position shifts of their ground states and ionic \bo curves [see Fig.~\ref{fig:BOcurves}(b,c)].
We investigate how it affects the photoemission dynamics in the present section by presenting and analyzing the orientation-resolved photoemission yields and stereo delays in these two model molecules, and addressing the relevance of fixed-nuclei simulations in this context. 

		\subsection{`Complete' simulations}
		
We used here the same pulse parameters as in the simulations with molecule \molA (Sec.~\ref{sec:A1phcomp}).

		\subsubsection{Signatures of electron-nuclei couplings}

A striking signature of the electron-nuclei couplings in the photoemission dynamics of \molC shows up in the delays inferred from the outgoing flux. In contrast with the previous case, the stereo delays measured according to Eqs.~\jref{eqn:tdv}--\jref{eqn:tau} strongly depend on the detection distance \rdt. This is illustrated in Table~\ref{tab:delC} with few representative cases. Not only it evidences a rather dramatic dependency of $\Delta\tau$  with respect to \rdt, well above the numerical accuracy of the simulations, but it turns out that $\Delta\tau$ monotonically diverges with increasing \rdt (not shown). This tells us that, the delays computed as such are irrelevant for an objective characterization of the ionization dynamics. 

\begin{figure}[t]
\includegraphics[width=0.8\figwidth]{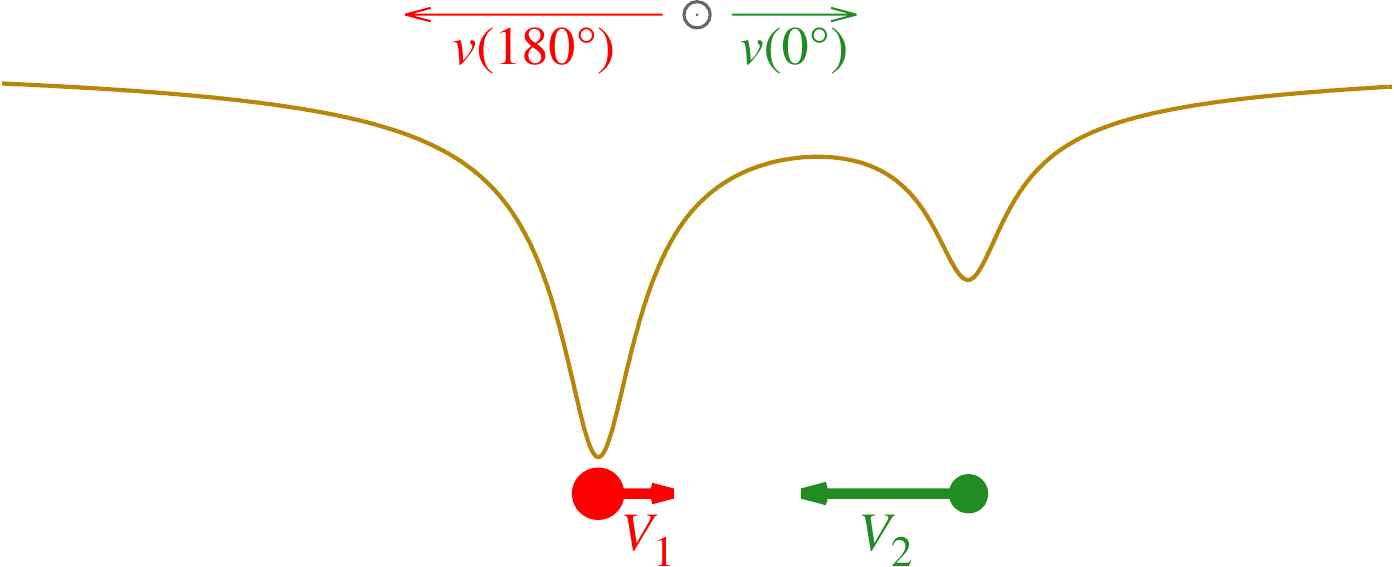}
\caption{\label{fig:Esharing} Sketch of the asymmetric average momentum sharing between the photoelectron (empty circle) and the nuclei (full circles) during the photoionization/early contraction of molecule \molC.  For the sake of clarity, the velocity asymmetries have been exagerated compared to the  ones obtained in the actual simulations.}
\end{figure}

Nevertheless their \rdt-dependency bear a crucial information: it suggests that the electron wave-packets travel on each side of the molecule with slightly different average velocities.
  This is the signature of an anisotropic momentum sharing between the nuclei and the active electron during the \textit{concerted} ionization/contraction of the molecular ion. It  involves the nuclear mass asymmetry since the lighter the nucleus, the more the photoelectron is prone to share kinetic energy with it. The principle, which applies to the average velocity, is sketched in Fig.~\ref{fig:Esharing}. Assuming that only the lighter nucleus (on the left) significantly moves, the photoelectron ends up with a smaller velocity if it exits in the direction of motion of that nucleus (towards the right) than in the other direction.
\begin{table}
\begin{tabular}{ccc}
\hline
\wX  & \rdt  & $\Delta\tau$  \\
\hline \hline
$32.6$ eV &\ 600 a.u.& 119 as \\
                 &\ 800  a.u.& 153 as \\
\hline
$34.1$ eV &\ 800  a.u.& \ 74 as \\
                 & 1000  a.u.& \ 89 as \\
\hline
\end{tabular}
\caption{\label{tab:delC} Stereo delay $\Delta\tau$ computed according to Eq.~\jref{eqn:tau} in molecule \molC for an illustrative set of detection distances \rdt and photon frequencies \wX. }
\end{table}
\begin{figure*}[t]
\includegraphics[width=2\figwidth]{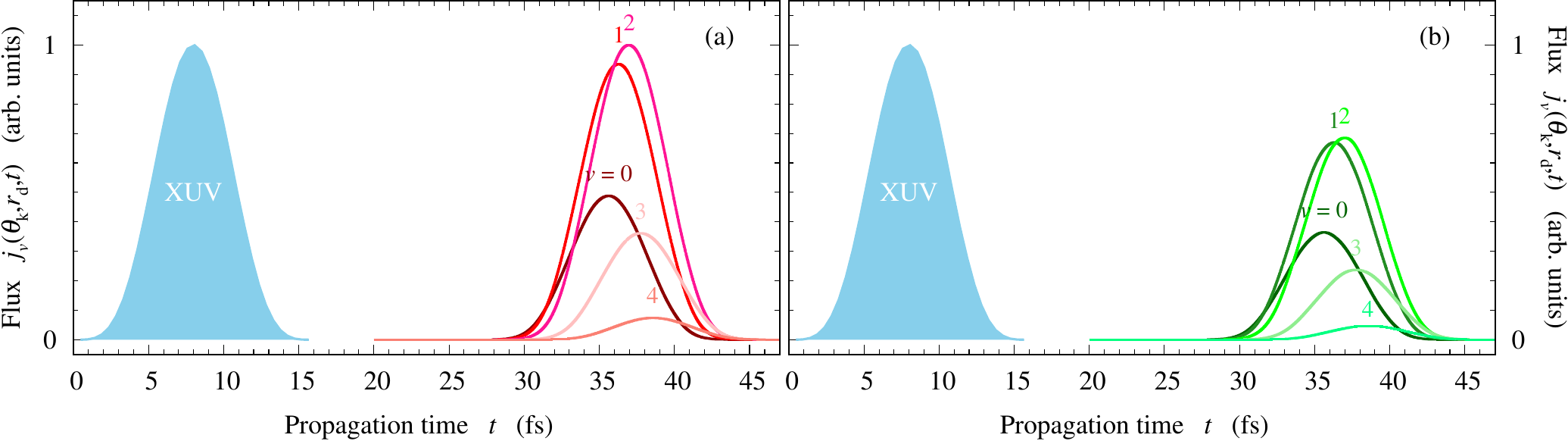}
\caption{\label{fig:flxC} Vibrationally resolved outgoing electron flux $j_v(\thetak,\rdt,t)$ computed, in the fully correlated simulations, at $\rdt=800$ a.u. on the left [$\thetak=180\degree$, frame (a)] and right [$\thetak=0\degree$, frame (b)] sides of  molecule \molC ionized by $\wX=35.67$ eV pulse, as a function of propagation time $t$, for the 5 first dominant channels. Each curve is labelled according to its corresponding channel ($v=0-4$). The flux are normalized to 1 at the overall maximum  reached in the  $v=2$ channel, on the left side. The temporal envelop of the ionizing pulse is represented by the blue filled curve.  \figOK
}
\end{figure*}

Properly accounting for this momentum sharing hence requires giving up global characterization and rather considering vibrationnally-resolved observables. In the present context, they are derived from the channel-resolved wave-packets
\begin{eqnarray}\label{eqn:phi_v}
\varphi_v(x,t)=\int\limits_0^\infty\,\chi_v(R)\Phi(x,R,t)\,\dd R
\end{eqnarray}
 in the vibronic \tdse simulations.
In Eq.~\jref{eqn:phi_v}, $\chi_v(R)$ are the vibrational eigenstates of the molecular ionic core hamiltonian \Hnu [see Eq.~\jref{eqn:H0}]. The individual channel functions are relevant as soon as they are uncoupled from each other, \ie in the asymptotic $x$ region where \Vne no longer depends significantly on $R$ (typically beyond few 100 a.u.). 
Each channel is assigned a specific ionization  potential 
\begin{eqnarray}\label{eqn:Ipv}
\Ip(v)&=&\E^{(+)}_v-\E^{(0)}_0,
\end{eqnarray}
where $\E^{(+)}_v$ is the energy of the corresponding ionization threshold (in the \bo framework, that is the eigenvalue of \Hnu associated with  $\chi_v$) and $\E^{(0)}_0$ the vibronic ground energy of the neutral molecule.

The key time-dependent observable becomes  the {\em vibrationally-resolved} electron flux
\begin{eqnarray} 
j_{v}(\thetak,\rdt,t)&=&\cos\thetak\times\\
&&\mathfrak{Im}\left\{
\varphi^\star_{v}(\rdt\cos\thetak,t) \varphi'_{v}(\rdt\cos\thetak,t)
\right\} \nonumber
\end{eqnarray}
computed at a distance \rdt on each side of the molecule. Vibrationnally-resolved anisotropic yields and delays can be computed out of the flux $j_{v}(\thetak,\rdt,t)$ in a similar fashion than in the integrated case, see Section~\ref{sec:A1ph}. 

The flux computed at the distance $\rdt=800$ a.u. on the left (a) and  right (b) sides of the \molC molecule  submitted to a $\wX=35.67$ eV pulse are shown in Fig.~\ref{fig:flxC}, for the first few popupated channels. The flux maxima of the different channels are time-shifted from each other, which is a manifestation of the energy conservation law
\begin{eqnarray}\label{eqn:Econs}
\varepsilon=\wX-\Ip(v).
\end{eqnarray}
 The larger $\Ip(v)$, the lower the photoelectron energy $\varepsilon$ and the larger the time needed to reach the detector.  As for molecule \molA, the asymmetry of molecule \molC results in anisotropic ionization yields, visible when comparing the flux magnitudes on the left and on the right. Similar results were obtained with molecule \molB (not shown).

\begin{figure}[t]
\includegraphics[width=\figwidth]{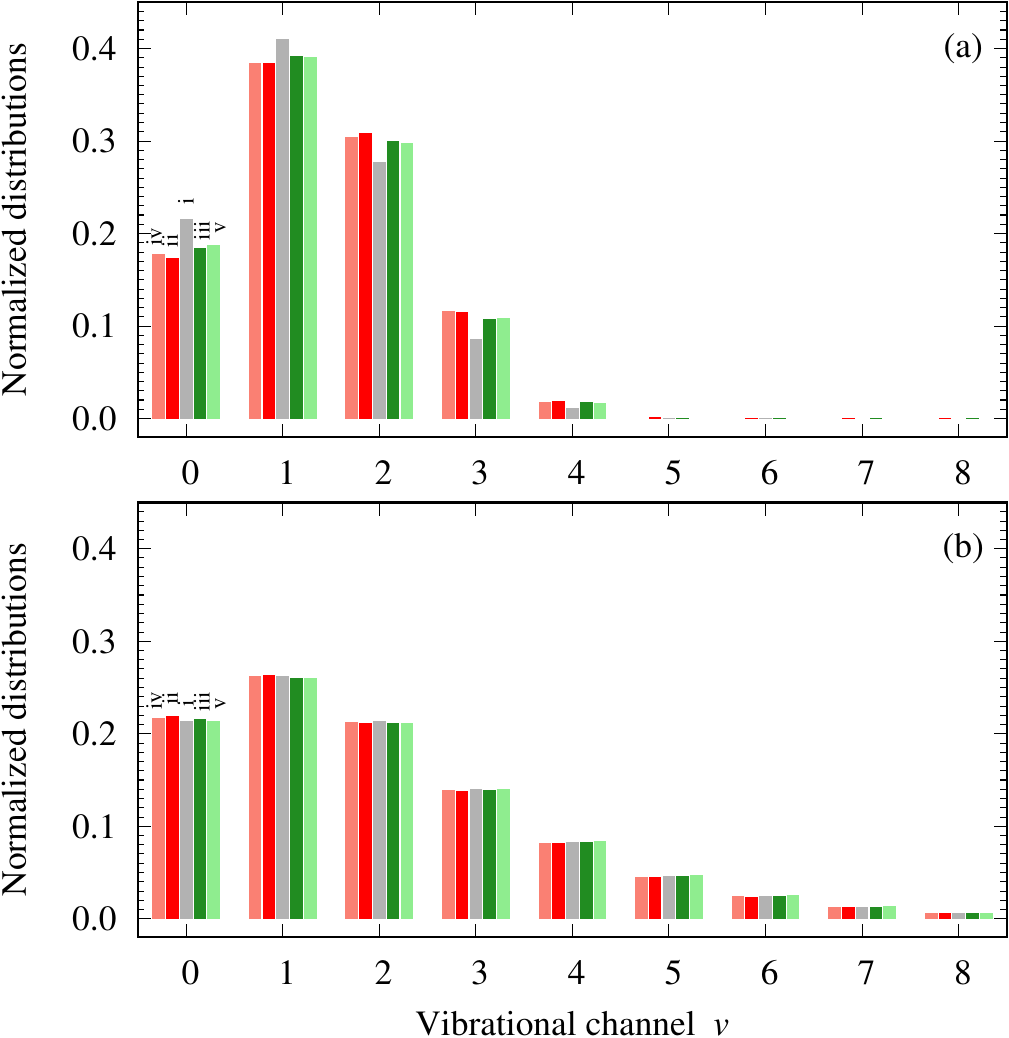}
\caption{\label{fig:FC} Vibrational distribution of photoemission from molecules \molC (a) and \molB (b) with a $\wX=35.67$ eV pulse. (i): Franck-Condon factors $F(v)$ [Eq.~\jref{eqn:FC}] ; (ii) and (iii): normalized yields computed in the fully correlated simulations, resp. $\Yld(v;180\degree)$ and $\Yld(v;0\degree)$ [Eq.~\jref{eqn:tdYRLv}]; (iv) and (v): normalized yields computed in the \bo framework at the optimal internuclear distances $\Ropt(v)$, resp. $\tilde \Yld(v;180\degree)$ and $\tilde  \Yld(v;0\degree)$ [Eq.~\jref{eqn:tiYRLv}]. 
The displayed data are normalized such that {\em each set} sums up to 1, for comparison purpose. 
}
\end{figure}

To get a better insight on the vibrational distributions upon ionization, we display in Fig.~\ref{fig:FC}, column i, the Franck-Condon (\FC) factors for molecules \molC (frame a) and \molB (frame b), defined as
\begin{eqnarray}\label{eqn:FC}
F(v)&=&\left\vert \int\limits_0^\infty\,\chi_v(R) \xi_0(R)\, \dd R\right\vert^2,
\end{eqnarray} 
where $ \xi_0(R)$ is the ground vibrational wave-function of the neutral molecule treated in the \bo framework. The \fc factors peak at $v=1$ for both model molecules, and extend significantly up to $v=4$ and $8$ for molecules \molC and \molB, respectively. They qualitatively reproduce the actual $v$-resolved yields 
\begin{eqnarray}\label{eqn:tdYRLv}
\Yld(v;\thetak)&=&\int\,j_v(\thetak,\rdt,t)\,\dd t,
\end{eqnarray}
computed on the left (columns ii) and right (columns iii) sides of the molecules ionized by the  illustrative $\wX=35.67$ eV pulse. 
 For the sake of comparison, the displayed data are normalized such that each set sums up to 1.

 \begin{figure*}[t]
\includegraphics[height=1.1\figwidth]{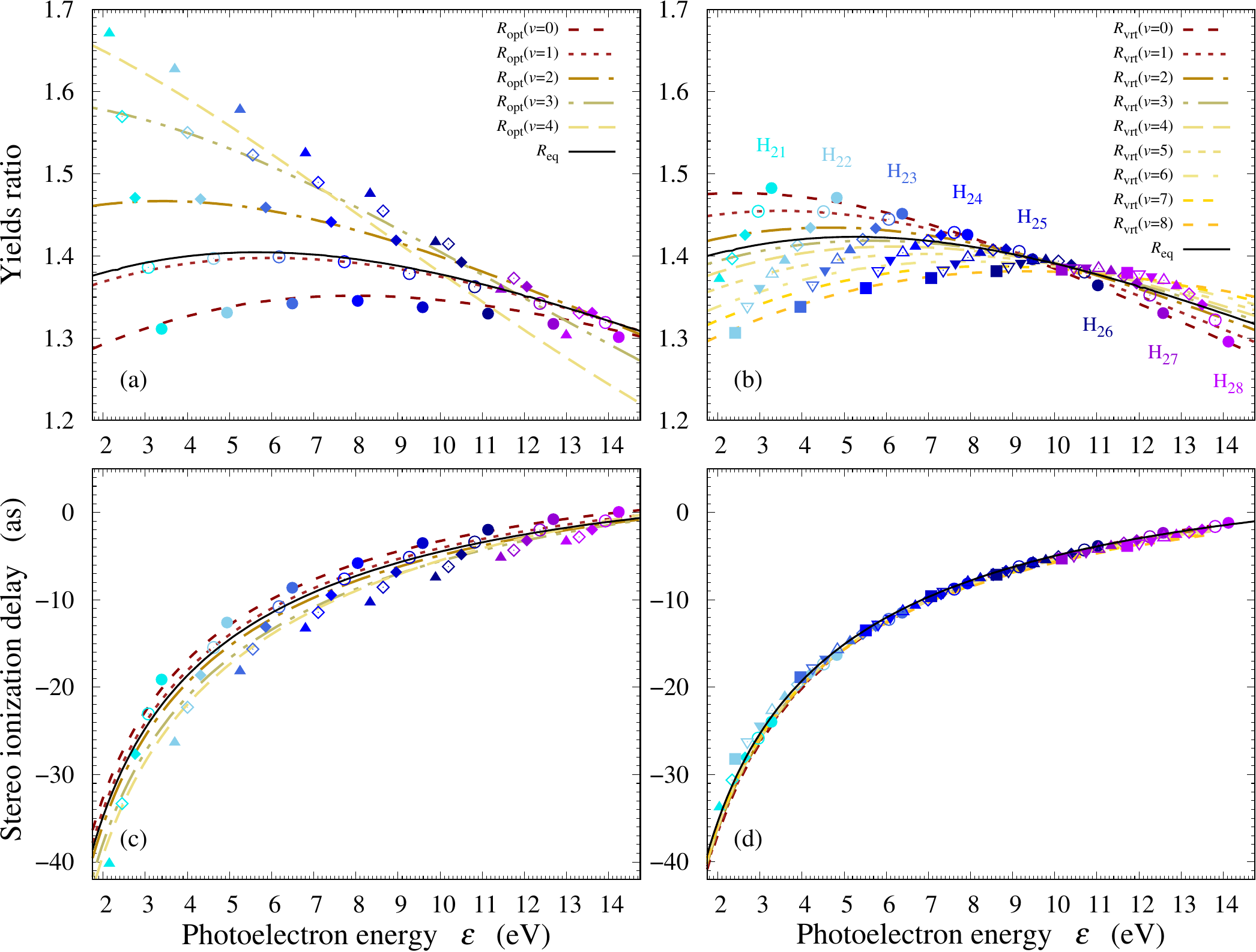}
\caption{\label{fig:ionBC} Vibrationnally resolved anisotropic photoemission from molecule \molC (left) and \molB (right). All data are plotted as functions of the photoelectron energy $\varepsilon$. Left/right  yields ratio (top) computed in the fully correlated simulations and in the \bo framework, resp. $\ratio(v)$ (symbols) and $\tilde \ratio(v)$ (lines); stereo delays (bottom) computed in  the fully correlated simulations and in the \bo framework, resp. $\Delta\tau(v)$ (symbols)  and $\Delta\wigdel(v)$ (lines). In all frames, the vibronic data are displayed for the first few significant vibrational channels: $v=0$ (full circles), $1$ (empty circles), $2$ (full diamonds), $3$ (empty diamonds), $4$ (full triangles up), $5$ (empty triangles up), $6$ (full triangles down), $7$ (empty triangles down) and $8$ (full squares). As specified by the inset key, the displayed \bo data were computed at a set of $v$-dependent optimal distances [$\Ropt(v)$] for molecule \molC, and of `vertical' distances [$\Rvrt(v)$] for molecule \molB (see text and Tab.~\ref{tab:RoptBC}). The \bo data  computed at the equilibrium distance ($\Req$) are also shown for both molecules.  \figOK
}
\end{figure*}

 A feature that cannot be included in the \fc factors alone is obviously the orientation dependency. 
In particular for molecule \molC, the vibrational distributions of the yields in Fig.~\ref{fig:FC}(a) display slight yet significant differences on the left and right emission sides. In spite of being small, these discrepancies are sufficient to result in different average photoelectron velocities, and to the above-mentioned \rdt dependance of the {\em integrated} left-right delay $\Delta\tau$.   We emphasize that the data in Fig.~\ref{fig:FC} are normalized to highlight the orientation-resolved vibrational {\em distributions}, and therefore discards the  vibrationally-resolved {\em magnitude} of the ionization probability on each side of the molecules.

		\subsubsection{Photoemission yields and delays}
		
A finer insight on the $v$-resolved photoemission anisotropy is provided in Fig.~\ref{fig:ionBC} which shows, for molecules \molC [frame (a)] and \molB [frame (b)], the yields ratio  
\begin{eqnarray}
\ratio(v)&=&\frac{\Yld(v;180\degree)}{\Yld(v;0\degree)}
\end{eqnarray} 
obtained with several values of \wX in the `complete' simulations (symbols), as a function of the photoelectron energy $\varepsilon$. Their values, comprised between $\sim1.2$ and $\sim1.7$, follow for each $v$ the global trend observed with molecule \molA, see Fig.~\ref{fig:ionA}(a). However, they display a clear additional $v$-dependency {for both molecules \molC and \molB}.

We  characterized the dynamics revealed in these simulations consistently with the unresolved case, see Eq.~\jref{eqn:tau}. Here, the channel-resolved stereo delays are defined as
\begin{eqnarray}\label{eqn:vtau}
\Delta\tau(v)&=&\tdv(v;180\degree)-\tdv(v;0\degree)-2\frac{\xini}{\sqrt{2\varepsilon}}
\end{eqnarray}
where the numerical \tof towards the virtual detector
\begin{eqnarray}\label{eqn:vtdv}
\tdv(v;\thetak)&=&\frac{\int\limits\, t\times j_v(\thetak,\rdt,t)  \,\dd t}{\int\limits\,  j_v(\thetak,\rdt,t)  \,\dd t},
\end{eqnarray}
is now computed in each $v$-channel.  In Eq.~\jref{eqn:vtau}, one should keep in mind  that $\varepsilon$ implicitly depends on $v$ through Eq.~\jref{eqn:Econs}, for a given \wX. We have notably checked that the \textit{vibrationally resolved} stereo delays $\Delta\tau(v)$ do not  depend on \rdt,  in contrast with the \textit{$v$-integrated} data commented before for molecule \molC (Table~\ref{tab:delC}). Indeed, for each vibrational channel, energy conservation ensures a symmetric asymptotic electron velocity. 
\begin{table*}
\begin{tabular}{|c|}
\multicolumn{1}{c}{\ }\\
\hline
$v$ \\
\hline
\Ip (eV)  \\ 
\hline
\Rvrt (\AA)  \\
\hline
\Ropt (\AA) \\
\hline
\end{tabular}
\begin{tabular}{
|ccccc|}
\multicolumn{5}{c}{(a)}\\
\hline
0 & 1 & 2 & 3 & 4 \\
\hline
29.17 & 29.48 & 29.78 & 30.10 & 30.41 \\ 
\hline
1.076 & 1.106 & 1.145 & 1.201 & 1.319 \\
\hline
1.084 & 1.113 & 1.146 & 1.203 & 1.248 \\
\hline
\end{tabular}
\begin{tabular}{|ccccccccc|}
\multicolumn{9}{c}{(b)}\\
\hline
 0 & 1 & 2 & 3 & 4 & 5 & 6 & 7 & 8 \\
\hline
 29.30 & 29.61 & 29.92 & 30.23 & 30.53 & 30.83 & 31.13 & 31.42 & 31.71 \\
\hline 
 1.160 & 1.135 & 1.115 & 1.098 & 1.082 & 1.066 & 1.055 & 1.041 & 1.031 \\
\hline
1.174 & 1.148 & 1.124 & 1.105 & 1.091 & 1.079 & 1.069 & 1.055 & 1.038 \\ 
\hline
\end{tabular}
\caption{\label{tab:RoptBC}
Channel-dependent ionization energy $\Ip$ [Eq.~\jref{eqn:Ipv}] and representative internuclear distances \Rvrt  and \Ropt [see text and Eq.~\jref{eqn:Rvrt}] for molecules  \molC (a) and \molB (b) in the main vibrational ionization channels $v$.
}
\end{table*}
\begin{figure*}[t]
\includegraphics[width=2\figwidth]{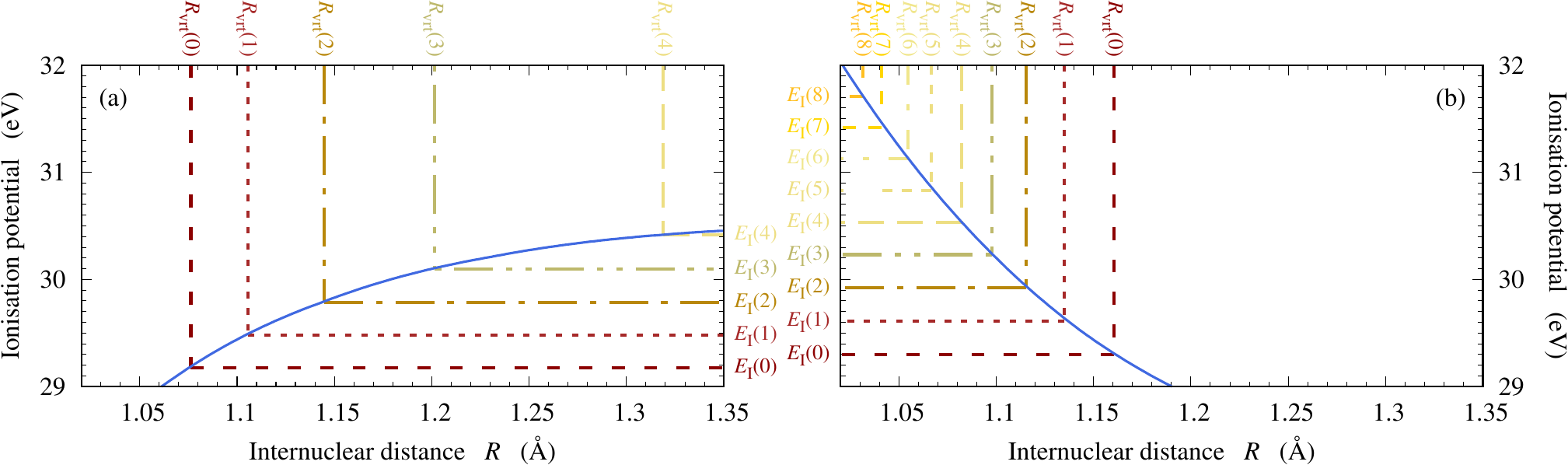}
\caption{\label{fig:RoptBC} Vertical ionization potential $\tilde\Ip(R)$ of molecules \molC (a) and \molB (b) as a function of $R$ (blue full curve), used to identify the internuclear distances $\Rvrt(v)$ fulfilling Eq.~\jref{eqn:Rvrt} out of the $v$-dependent ionization potentials $\Ip(v)$. \figOK 
}
\end{figure*}

The obtained $v$-dependent stereo delays are plotted as symbols in Fig.~\ref{fig:ionBC} for molecules \molC [frame (c)] and \molB [frame (d)]. We first note that they globally behave like the stereo delays reported  for molecule \molA in Fig.~\ref{fig:ionA}(b), with a comparable order of magnitude which decays smoothly towards 0 when the photoelectron energy $\varepsilon$ increases. Besides, a marked   $v$-dependency is observed in molecule \molC, reminiscent of the behavior of the corresponding yields ratios [frame (a)]. In molecule \molB however, the values of $\Delta\tdv(v)$ appear to  depend mostly on $\varepsilon$ regardless of the vibrationnal channel, since all of them follow a common spectral evolution. This is consistent with the normalized vibrational distributions displayed in Fig.~\ref{fig:FC}(b), which are almost identical in the two directions, as pointed out before.

		\subsection{Simulations at fixed internuclear distances}\label{sec:Ropt}

In this section, we investigate the possibilities to retrieve the orientation-dependent data, including their $v$-dependencies, by analyzing electronic continuum wave-functions obtained in simulations with fixed nuclei. 

The yields ratio and delays computed for both molecules in the \bo framework at \Req according to Eq.~\jref{eqn:tratio} are displayed as a full black lines in  Fig.~\ref{fig:ionBC}. They lie near the data obtained for the dominant channels  in the full \tdse simulations (between $v=1$ and $2$), and very close to the equivalent results obtained with \molA [see Fig.~\ref{fig:ionA}(a)]. Nevertheless, and quite obviously, simulations performed at a \textit{single} fixed internuclar distance cannot reproduce the non-trivial $v$-dependency observed in the full simulations for the yields ratio and stereo delays of molecule \molC, and for the yields ratio of molecule \molB.

Below, we use the simulations performed on molecule \molC to identify and interprete $v$-dependent geometries that allow reproducing the results of the vibronic simulations with the fixed nuclei approach. We then assess our interpretation by applying it to fixed nuclei simulations on molecule \molB.

		\subsubsection{Optimal conformations for molecule \molC}

By scanning through  the support of the initial vibrational state $\xi_0(R)$ of molecule \molC ($R\sim1.0 - 1.3$ \AA), we identified empirically a set of channel-specific optimal distances $\Ropt(v)$ for which both the yields ratio and the stereo delays in the \bo approach reproduce the ones of the complete vibronic simulations. The  values of $\Ropt(v)$  obtained are reported in Tab.~\ref{tab:RoptBC}(a). The orientation- and channel-resolved yields, now computed in the \bo framework at a given optimal conformation for each channel, 
\begin{eqnarray}\label{eqn:tiYRLv}
\tilde \Yld(v;\thetak) &=& \vert A(\Ropt(v);\thetak)\vert^2\times F(v), 
\end{eqnarray}
are displayed for the left and right sides  in columns iv and v  of Fig.~\ref{fig:FC}(a), respectively. 
They reproduce very well the asymmetric $v$-dependency of the actual yields $\Yld(v;\thetak)$ (columns ii and iii). The corresponding yields ratio $\tilde\ratio(v)$ and stereo delays $\Delta\wigdel(v)$, extracted from the \scwf computed at the optimal conformation for each $v$ are displayed as full lines in frames (a) and (c) of Fig.~\ref{fig:ionBC}. They are in excellent agreement with the $v$-dependent data obtained in the fully correlated simulations for both molecules, apart from the $v=4$, as will be discussed below.
 
%

\subsubsection{Physical interpretation}

The optimal molecular conformations can be interpreted by looking at the  $R$-dependent ionization potential $\tilde\Ip(R)$  [Eq.~\jref{eqn:IpR}], which for molecule \molC increases monotonically with $R$ within the support of $\xi_0(R)$, see Fig.~\ref{fig:RoptBC}(a). On the same figure, we indicate the  internuclear distances  $\Rvrt(v)$ at which $\tilde\Ip$ matches the $v$-dependent ionization potential [Eq.~\jref{eqn:Ipv}],
\begin{eqnarray}\label{eqn:Rvrt}
\tilde\Ip[\Rvrt(v)]&=&\Ip(v).
\end{eqnarray}
These ``vertical'' internuclear distances are reported in Table~\ref{tab:RoptBC}(a), for comparison with the optimal distances found empirically. It turns out that \Ropt matches \Rvrt within $\sim 1 \%$ in all the considered channels, except $v=4$  ($\sim 5\%$).

We repeated the same procedure for molecule \molB. The optimal internuclear distances, $\Ropt(v)$, that we found empirically are reported in Table~\ref{tab:RoptBC}(b), together with the `vertical' ones, $\Rvrt(v)$, extracted from the ionization potential displayed in Fig.~\ref{fig:RoptBC}(b). It is here also a monotonic function of $R$, with opposite variations than for molecule \molC (as mentioned earlier, molecule \molC expands upon ionization, while molecule \molB contracts upon ionization).  The \BO data displayed as lines in Fig.~\ref{fig:ionBC}(b) and (d) for molecule \molB were directly obtained with $\Rvrt(v)$, for each channel. The differences with the data obtained using $\Ropt(v)$, not shown, would be hardly discernible on that figure. The agreement between the `complete' simulations and the fixed-nuclei approach is here excellent in all considered channels -- including the ionization ratios [frame (b)] which follow spectral trends that significantly depend on $v$.

These results provide a simple physical criterion for selecting \textit{a priori} a set of optimal conformations which allow reproducing, in a fixed-nuclei framework, the details of $v$-dependent molecular photoionization dynamics. It is nevertheless predictable that these specific distances can be representative of the $v$-dependent photoemission dynamics only when  $\tilde\Ip(R)$ varies monotonically with $R$. In particular for molecule \molC the $\tilde\Ip(R)$ variations are less pronounced as $R$ grows towards the $\Rvrt$ of the highest channels, within the studied range. Consistently, the data computed in the \bo framework for molecule \molC turned out to be less sensitive to small $R$ variations beyond $\Ropt(v=3)$ -- to the extent that we could not find an optimal distance for the $v=4$  channel with the same accuracy as for other channels. This is particularly visible on the yields ratio [Fig.~\ref{fig:ionBC}(a)], where the \bo data never reach the `exact' ones obtained for the highest ($v=4$) channel, the displayed \bo data corresponding to $\Ropt(v=4)$ being the closest one could get. This however concerns a minor channel [see Fig.~\ref{fig:FC}(a)]. For molecule \molB, the narrower dispersions of the data  can be directly related to the sharp  monotonic decay of $\tilde\Ip$ when $R$ is increased, which implies a narrow dispersion of $\Rvrt(v)$.

 	\subsection{Illustrative applications of the optimal conformations}\label{sec:applications}

 In this last section, we address the relevance of the optimal conformations beyond the context in which they were identified. We  investigate their relevance first in simulations of interferometric  \rabbit measurements, and then to include electron-ion coherences during broadband 1-photon ionization
 
 The \rabbit scheme, together with the attosecond streaking approach, provide indirect ways to access photoemission dynamics\footnote{ See~\cite{berkane2024a} and references therein for discussions dedicated to the links between \rabbit measurements and fundamental photoemission delays.}, the direct time domain approaches used in our simulations having no experimental equivalent with attosecond resolution (\eg  magnetic bottles have typical deadtimes of few 10 ns~\cite{palaudoux2008a} and time-to-digit convertor resolutions in the 100 ps range~\cite{penent2020a}). 
 
 These are interferometric techniques, where the time domain information is extracted from the spectral variations of measured phases. Coherence is thus an essential issue in the design and exploitation of these approaches, the purpose of which is to reveal the fundamental dynamics of essentially quantum processes.
  
 \subsubsection{Intra-channel coherences: \rabbit interferometry}
 \begin{figure}[t]
\includegraphics[width=\figwidth]{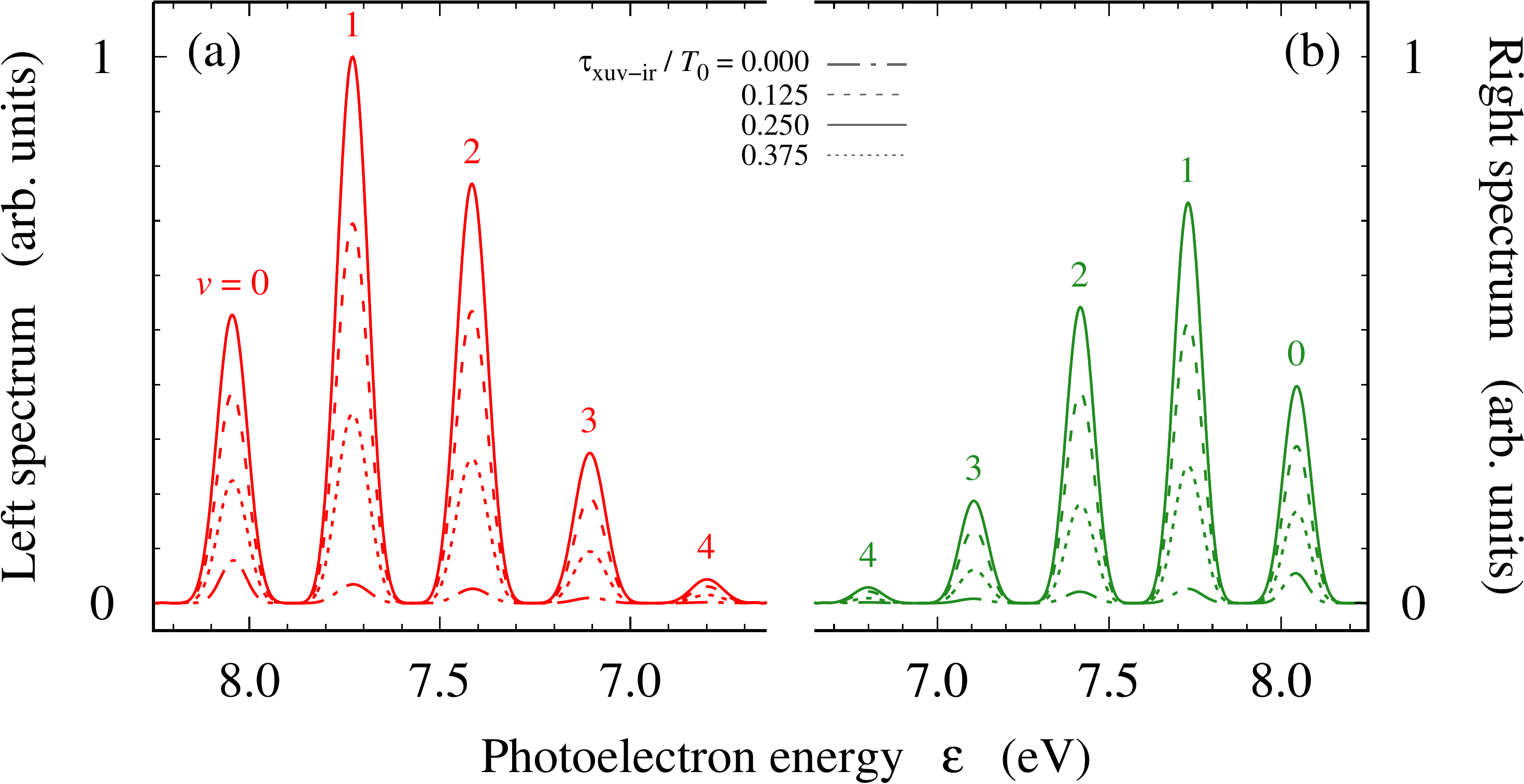}
\caption{\label{fig:spcrabC} Photoelectron spectra recorded in \rabbit simulations, on the left (a) and right (b) sides of molecule \molC. The spectral range is focused on sideband 24, and each peak is associated with a specific ionization channel $v=0-4$ (see peak labels). The different spectra were obtained with different values of the pump-probe delay $\tauxuvir$ (see inset keys, where $\TIR=2\times\pi/\wIR$ is the laser period). 
}
\end{figure}

\begin{figure*}[t]
\includegraphics[height=1.075\figwidth]{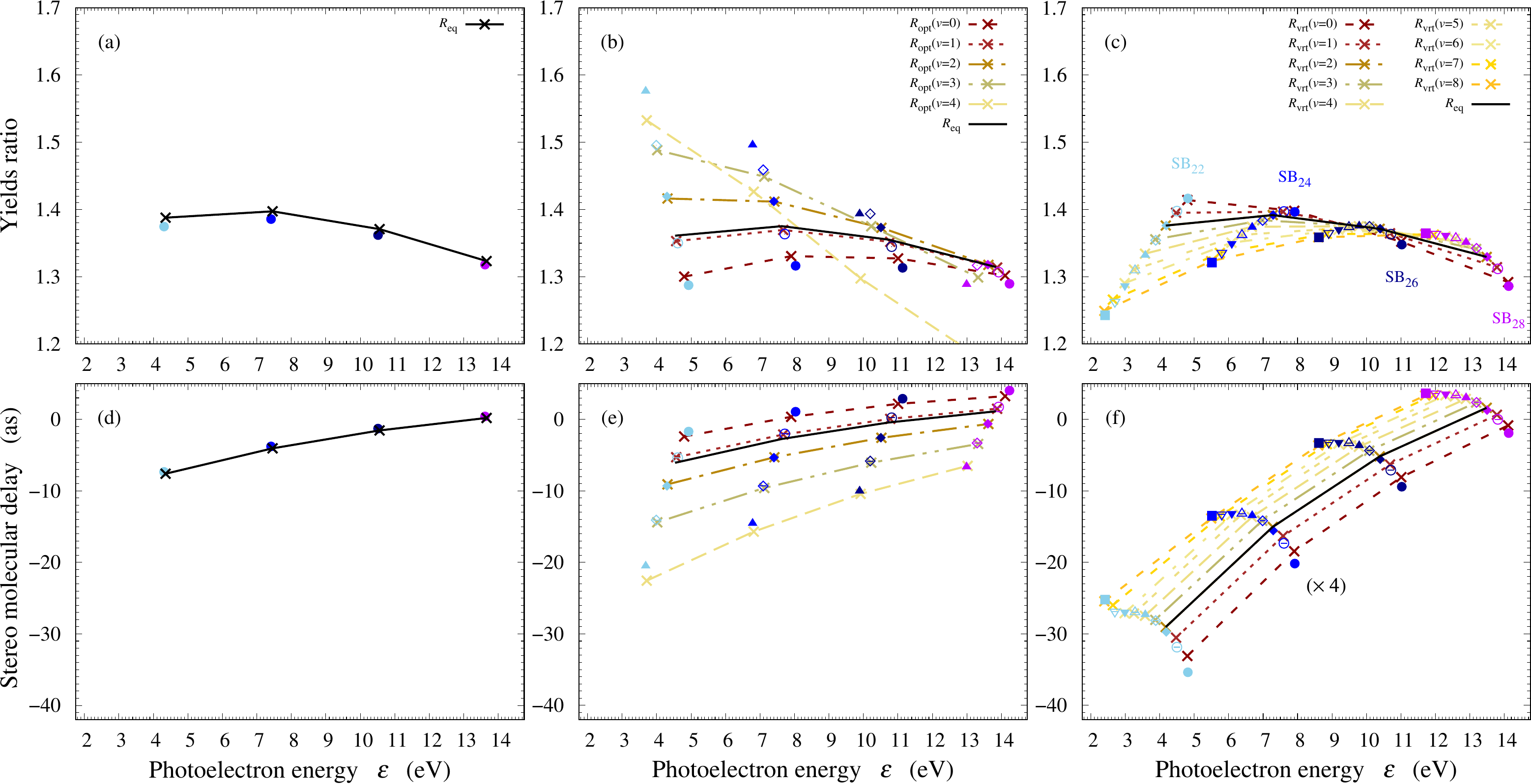}
\caption{\label{fig:rabABC}  Vibrationnally resolved anisotropic \rabbit simulations in molecule \molA (left), \molC (center) and \molB (right). Data derived from the analysis of sidebands 22 to 28 (right to left) plotted as functions of the photoelectron energy $\varepsilon$. Left-right  yields ratio (top) computed in the fully correlated simulations and in the \bo framework, resp. $\ratiorab(v)$ (symbols) and $\tratiorab(v)$ (crosses+guidelines); stereo molecular delays (bottom) computed in  the fully correlated simulations and in the \bo framework, resp. $\Delta\rabdel(v)$ (symbols) and $\Delta\tilde\tau_{\mbox{\tiny mol}}(v)$ (crosses+guideline). In each frame, the data are displayed for the first few significant vibrational channels: $v=0$ (full circles), $1$ (empty circles), $2$ (full diamonds), $3$ (empty diamonds), $4$ (full triangles up), $5$ (empty triangles up), $6$ (full triangles down), $7$ (empty triangles down) and $8$ (full squares). Consistently with the results shown in Fig.~\ref{fig:ionBC}, the displayed \bo data were obtained with the set of $v$-dependent optimal distances [$\Ropt(v)$] identified in 1-photon simulations for molecule \molC, and of `vertical' distances [$\Rvrt(v)$] for molecule \molB (see text and Tab.~\ref{tab:RoptBC}). They were also computed at the equilibrium distance ($\Req$) for all molecules. The data displayed in frame (f) are all multiplied by 4 for a better readability.
}
\end{figure*}

Following a standard 800-nm \rabbit scheme~\cite{paul2001a,muller2002a}, we simulated photoemission from the three model molecules in presence of the fundamental \ir field and a comb of its odd harmonics \HA{q} in the \xuv domain (orders $q$ from $21$ to $29$), with an adjustable pump-probe delay \tauxuvir.  We computed vibrationally- and orientation-resolved photoelectron spectra 
\begin{eqnarray}
\sigma(v;\varepsilon,\thetak)&=&\vert \tda(v;\varepsilon,\thetak)\vert^2
\end{eqnarray}
out of the final amplitudes corresponding to the ionized molecule, $\tda(v;\varepsilon,\thetak)$. The latter were obtained in the time-dependent simulations through a position-to-momentum Fourier transform of the channel wave packets $\varphi_v(x,t)$ accumulated in the asymptotic region at each time $t$ of the propagation (see Appendix C of~\cite{caillat2005a}). 

As expected, we obtained sideband (\SB{}) peaks (orders $22$ to $28$) resulting from 2-photon $\xuv\pm\ir$ transitions, the intensities of which oscillate when \tauxuvir is tuned~\cite{veniard1996a}. We used long enough pulses (80 fs for the \xuv and for the \ir) to resolve the vibrational channels in each sideband. This is illustrated in Fig.~\ref{fig:spcrabC}, which shows the vibrationally-structured, orientation and delay-dependent \SB{24}  obtained with molecule \molC. Similarly to the 1-photon case, the relative peak intensities evidence a significantly larger photoemission probability towards the left (a) than towards the right (b). Note however that the orientation-dependency of the \SB{} oscillations are too subtle to be resolved visually in this figure. This is consistent with the ultrashort timescale of the simulated non-resonant photoemission processes~\cite{berkane2024a}.   
 
We thus performed a \rabbit analysis consisting in fitting the generic function given in Eq.~8 of~\cite{berkane2024a} to the \tauxuvir evolution of each sideband peak associated with a given vibrational channel $v$, and a given direction \thetak. The fitting procedure gives access to the orientation and channel-resolved phases of the peak oscillations $\vartheta(v;\thetak)$, as well as to the \tauxuvir-averaged photoelectron yields $P(v;\thetak)$.
Since the \XUV components in our simulations were all synchronized (\ie they carry no attochirp), $\vartheta(v;\thetak)$ directly corresponds to the so-called ``molecular phase'', see~\cite{berkane2024a} and references therein.

We performed simulations both in the fully correlated approach, and in the \bo framework. The main  results are shown in Fig.~\ref{fig:rabABC}, for \SB{22}  to \SB{28} in molecules \molA (left), \molC (center) and \molB (right). The disconnected symbols in the upper and lower frames  respectively correspond to the ratio of the \tauxuvir-averaged yields, 
\begin{eqnarray}
\ratiorab(v)&=&\frac{P(v;180\degree)}{P(v;0\degree)}
\end{eqnarray}
and the ``stereo molecular delay''~\cite{klunder2011a,chacon2014a} 
\begin{eqnarray}
\Delta\rabdel(v)&=&\frac{\vartheta(v;180\degree)}{2\times\wIR}-\frac{\vartheta(v;0\degree)}{2\times\wIR}
\end{eqnarray}
 obtained in the fully correlated approach for the main $v$  channels. As in the 1-photon case, these quantities follow standard general trends which can be summarized by looking at the single-channel results of molecule \molA. Its yields ratio $\ratiorab(0)$ decays slowly, from $\sim1.4$ down to $\sim1.3$ over the covered energy range, which is reminiscent of the 1-photon counterpart shown in Fig.~\ref{fig:ionA}(a). The stereo molecular delay $\Delta\rabdel(0)$ decays in magnitude, from $-10$ as at \SB{22} down to near $0$ as at \SB{28}. The relationship between the latter and  the 1-photon ionization delay is the subject of the companion paper~\cite{berkane2024a} and will not be further discussed here. Beyond the trends commented above, the results for molecule \molC and \molB displays  clear $v$-dependencies, both in the yields ratios and in the stereo molecular delays, that are somehow more pronounced than in the 1-photon case.
 
 The crosses with linear guidelines correspond to the equivalent quantities, resp. $\tratiorab(v)$ and $\Delta\tilde\tau_{\mbox{\tiny mol}}(v)$, obtained in time-dependent simulations at {\em fixed} internuclear distances. For molecule \molA, we used again the equilibrium distance \Req, which unsurprislingly perfectly reproduces the results of the `complete' simulations. Note that these fixed nuclei data appear in Fig. 6 of ~\cite{berkane2024a}. For molecule \molC, we additionally show the BO results obtained with the $v$-dependent optimal distances \Ropt identified in the 1-photon simulations (see Table~\ref{tab:RoptBC}), while for molecule \molB we used the vertical internuclear distances $\Rvrt(v)$. The data obtained at \Req both with molecules \molC and \molB are similar to the ones obtained with molecule \molA. Regarding the results provided at  optimal conformations, one clearly sees that the \bo framework fails to reproduce the yields ratio for molecule \molC in the minority $v=4$ channel [frame (b)], already discussed in the 1-photon case. Apart from this, the data obtained in the \bo framework, including the stereo molecular delays in all channels for both molecules are in excellent agreement with the ones obtained in the `complete' simulations. This illustrates the capacities of \bo approaches with appropriate internuclear distances to simulate vibrationnally-resolved \rabbit  measurements with high fidelity.   
 		\subsubsection{Inter-channel coherences: broadband photoemission}
\begin{figure}[t]
\includegraphics[width=\figwidth]{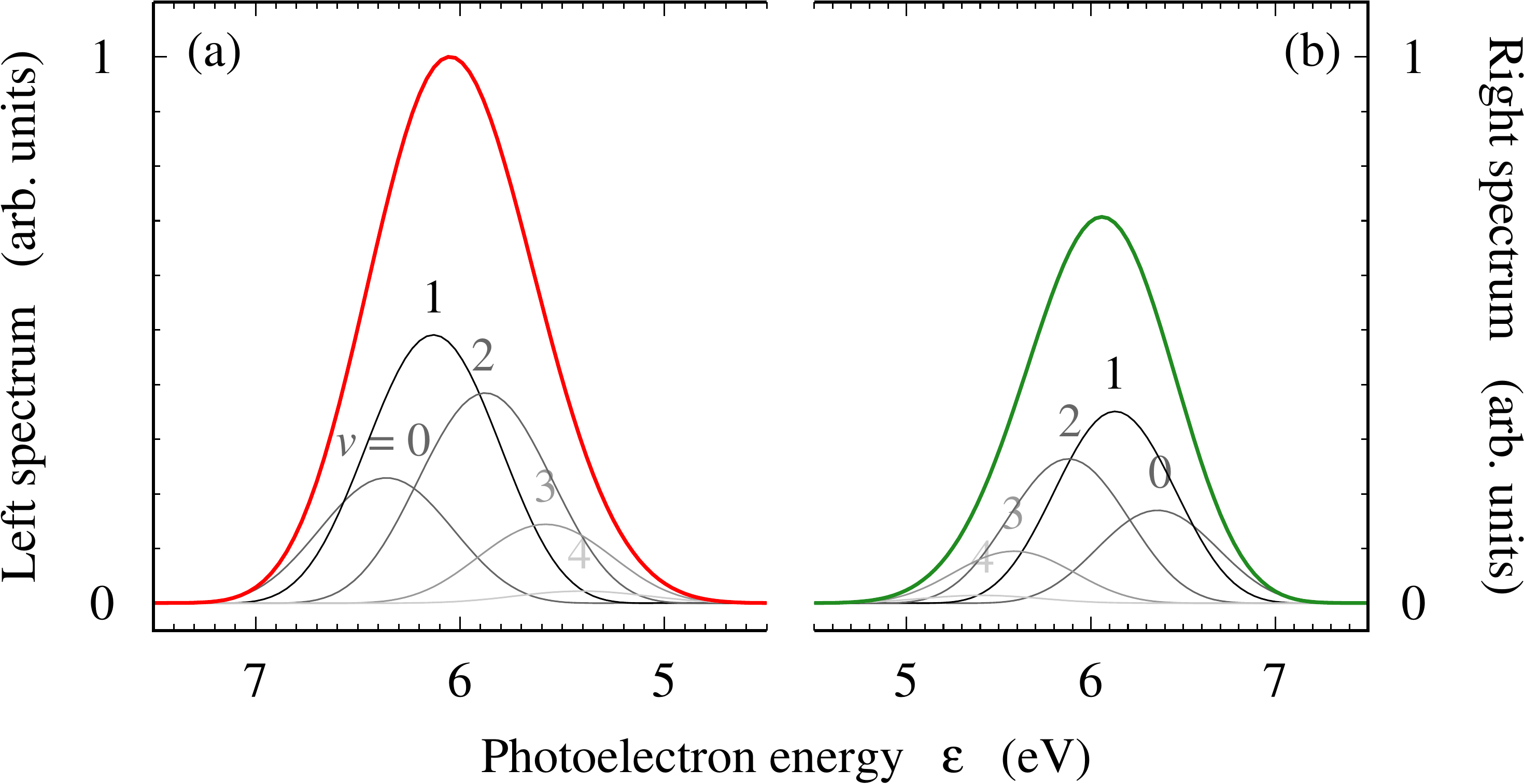}
\caption{\label{fig:spcrdmC} Photoelectron spectrum recorded on the left (a) and right (b) sides of  molecule \molC, upon single-photon ionization with a 40 fs light pulse of $37.23$ eV central frequency (\HA{24} of a 800 nm field). Thin gray lines: $v$-resolved spectra ($v=0-4$, see peak labels); thick coloured lines: $v$-integrated spectra. \figOK}
\end{figure}
\begin{figure*}[t]
\includegraphics[width=1.5\figwidth]{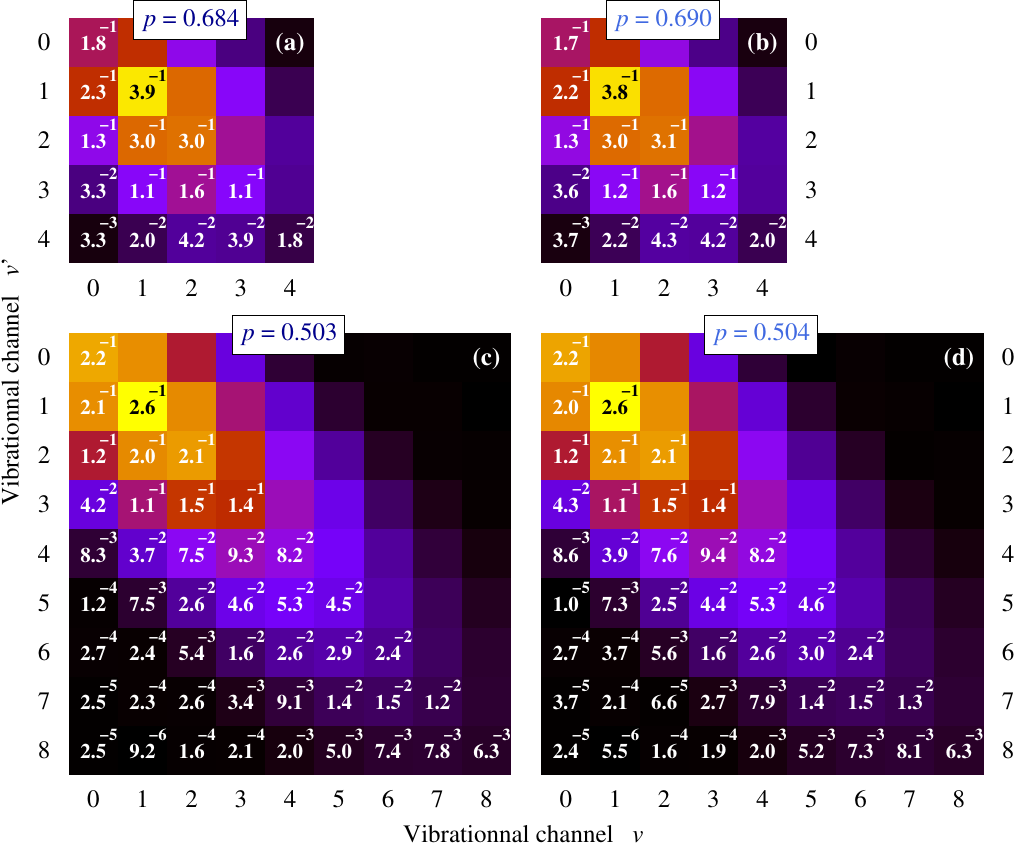}
\caption{\label{fig:rdm} Photoionization of molecule \molC (top) and \molB (bottom) with a broadband \xuv pulse (same simulations as Fig.~\ref{fig:spcrdmC}). Modulus of the final ion's reduced density matrix $\rho\ion(v,v')$ [Eq.~\eqref{eqn:rhoion}] in the complete simulations (left) and in simulations at fixed conformations $\Rvrt(v)$ (right). The values of the matrix elements' modulii are provided in the lower triangles (the powers of ten are given in superscript). The matrices are displayed for the main open channels. Each one is normalized to have a trace equal to 1, consistently with the data displayed in Fig.~\ref{fig:FC}. The value of the purity $p$ [Eq.~\eqref{eqn:purity}] is indicated at the top of each matrix. }
\end{figure*}

As a last case, we now highlight the capacities of fixed-nuclei approaches to keep track of {\em inter}-channel coherences, when reconstructing the complete vibronic wave-packet out of a set of simulations performed at the $v$-dependent optimal conformations. To this end, we simulated the photoionization of molecules \molC and \molB with a broad \xuv pulse of central frequency  $\wX=37.23$~eV (\HA{24} of a $800$ nm laser) and a duration of $8$ fs \fwhm (3 fundamental \ir cycles), in the 1-photon perturbative regime. The bandwidth of this pulse overlaps few vibrationnal channels in the photoelectron spectra, as shown in Fig.~\ref{fig:spcrdmC} for molecule \molC. One therefore partially looses the vibrationnal resolution when looking at the $v$-integrated spectra, in contrast to the results obtained with narrower pulses, see for instance Fig.~\ref{fig:spcrabC}.

From the final channel-selective amplitudes in the continuum $\tda(v;\varepsilon,\thetak)$, we computed the reduced density matrix (\rdm) of the ion in the final state as
\begin{eqnarray}\nonumber
\rho\ion(v,v')&=&\sum\limits_{\thetak=0\degree,180\degree}\int\limits_{\Delta}\,d\varepsilon\, \left[\tda(v;\varepsilon,\thetak)\right]^\star \tda(v';\varepsilon,\thetak) \\ \label{eqn:rhoion}
\end{eqnarray}
where $[\ ]^\star$ denotes the complex conjugate and the spectral integration range $\Delta$ is restricted to the overall support of the photoelectron spectrum. The modulus of the \rdm computed in the complete simulations for molecules \molC and \molB  in the main vibrationnal channels are respectively shown in frames (a) and (c) of Fig.~\ref{fig:rdm}. For both molecules, we observe non vanishing off-diagonal elements which are signatures of coherences between overlapping channels ($\vert v'-v \vert \lesssim 3$). 

In the fixed nuclei simulations, we reconstructed the final \rdm of the ion out of the amplitudes $\tilde\tda(\Ropt(v);\varepsilon,\thetak)$ obtained in time-dependent simulations using the $v$-dependent vertical internuclear distances $\Rvrt(v)$ for both molecules. The results are shown in frames (b) and (d) of Fig.~\ref{fig:rdm} for molecules \molC and \molB respectively. The overall agreement with the complete simulations is excellent both for the populations (diagonal elements) and the coherences.  
To further evaluate the quality of these reconstructed vibrational \rdm, we computed in each simulation the purity of the final state 
\begin{eqnarray}\label{eqn:purity}
p=\frac{\trace{\rho\ion^2}}{\trace{\rho\ion}^2},
\end{eqnarray} 
 where \trace{} denotes the trace application. The purity values are indicated in Fig.~\ref{fig:rdm} at the top of each matrix\footnote{We verified that the normalized matrices and the purities remain practically unchanged when looking at the $\theta$-dependent \RDM obtained as in Eq.~\eqref{eqn:rhoion} but without angular integration (results not shown).}. The purity in fixed-nuclei results accurately match the `exact' ones, within less than $1\%$, for both molecules.  It is worth noting that, when using the $\Ropt(v)$ conformations rather than the $\Rvrt(v)$ ones, the fixed-nuclei simulations agree with the `exact' ones with a $\sim10\%$ accuracy only (results not shown). On the one hand, this underlines the sensitivity of the computed purity with respect to the conformations used in the fixed nuclei approach. On the other hand, it emphasizes the remarkable efficiency of the combined vertical conformations to reproduce the exact interchannel coherent dynamics.

\section{Summary and conclusion}\label{sec:conclusion}

We studied numerically how nuclear motion affects the dynamics of orientation-resolved  photoemission in asymmetric diatomic model molecules presenting minimal vibronic couplings. We considered near-threshold photoemission in absence of any resonance, where the so-called stereo Wigner delays amount to few tens of attoseconds or less.

We have shown that the intrinsic molecular asymmetry results in an anisotropic electron-ion momentum sharing which, as slight as it may be, may prevent us from assigning unambiguously a stereo Wigner delay to the {\em channel-averaged} photoemission process.  Indeed, a small asymmetry in the average photoelectron energy leads to stereo delay values that diverge when the (virtual) detection distance increases. This is  circumvented by considering vibrationnally-resolved photoemission. However, on the theory side, comprehensive time-dependent vibronic simulations of molecules interacting with external fields are restricted to smaller molecules such as H$_2$ with a limited range of physical and numerical parameters, or to simplified low dimensional model molecules such as the ones used in the present work.

Therefore, we investigated ways of retrieving the vibrationnally-resolved photoemission dynamics revealed by the `complete' vibronic simulations, out of more standard and broadly applicable time-independent approaches with fixed nuclei. We  found empirically that each vibrationnal channel could be assigned an effective internuclear distance that reproduces the channel-resolved orientation-dependent photoemission yields and delays with a good accuracy. Furthermore, we identified a physical criterion, relying on the molecule's ionization energies, that allows selecting {\em a priori} the $v$-dependent effective internuclear distances:  It corresponds to the distance for which the vertical ionization potential matches the exact channel-dependent ionization potential. Identifying such unique effective molecular conformations is expected to work efficiently as long as the vertical ionization potential varies significantly when the molecular conformation is changed. Retrieving photoemission dynamics with vibrational resolution out of fixed nuclei simulation is also expected to work as long as photoemission takes place with little vibronic correlation, typically in smooth continua with sufficiently separated ionic vibrational levels. 

Eventually, we  assessed the relevence of the fixed-nuclei approach beyond the context in which we identified the effective internuclear distances. We first showed that it could be used to accurately simulate anisotropic vibrationnally-resolved \rabbit interferometry, both in terms of phase and amplitudes. Then, we investigated  the fixed nuclei approach capacities to account for interchannel coherences in broadband photoionization of our model molecules, where the different channels significantly overlap. We showed that the fixed-nuclei approach could satisfactorily reproduce both the ion's reduced density matrix and  purity in the final state. This approach is thus of particular interest to model attosecond resolved photoemission dynamics in  benchmark molecules without the need to resort to fully-fledged vibronic approaches. It could be applied to simulate  attosecond time-resolved interferometry,  highly non-linear processes such as strong field ionization~\cite{shvetsov-shilovski2023a} or molecular high-order harmonic generation~\cite{lein2005a,zair2013a,risoud2017a,zhang2019a,labeye2019a}, and to investigate ultrafast  decoherence processes in molecule~\cite{vrakking2021a,vrakking2022a,nabekawa2023a} of  crucial importance in attochemistry.

\begin{acknowledgments}
This research received the financial support of the French National Research Agency through Grants No. ANR-15-CE30-0001-CIMBAAD and ANR-20-CE30-0007-DECAP.
\end{acknowledgments}




%

\end{document}